\documentclass{pinchcr}   
\usepackage{bbold}

\usepackage{epsfig}
\usepackage{amsmath}
\usepackage{pstricks}
\usepackage{pst-coil}

\newcommand{\be}{\begin{equation}}   
\newcommand{\ba}{\begin{eqnarray}}
\newcommand{\ea}{\end{eqnarray}}

\newcommand{\rmO}{\mathrm{O}}
\newcommand{\rmR}{\mathrm{R}}
\newcommand{\rme}{\mathrm{e}}
\newcommand{\rmd}{\mathrm{d}}

\newcommand{\calA}{\mathcal{A}}
\newcommand{\calC}{\mathcal{C}}
\newcommand{\calD}{\mathcal{D}}
\newcommand{\calE}{\mathcal{E}}
\newcommand{\calF}{\mathcal{F}}

\newcommand{\calL}{\mathcal{L}}
\newcommand{\calO}{\mathcal{O}}
\newcommand{\calP}{\mathcal{P}}
\newcommand{\calR}{\mathcal{R}}
\newcommand{\calS}{\mathcal{S}}
\newcommand{\calV}{\mathcal{V}}

\newcommand{\bfk}{{\bf k}}
\newcommand{\bfx}{{\bf x}}
\newcommand{\bfy}{{\bf y}}
\newcommand{\bfz}{{\bf z}}

\newcommand{\tr}{\mathrm{tr}}

\newcommand{\psibar}{\overline{\psi}}
\newcommand{\chibar}{\overline{\chi}}

\newcommand{\udeg}{\overline{\mathrm{deg}}}
\newcommand{\ldeg}{\underline{\mathrm{deg}}}

\newcommand{\mq}{m_\mathrm{q}}
\newcommand{\muq}{\mu_\mathrm{q}}
\newcommand{\mc}{m_\mathrm{c}}

\newcommand{\msbar}{\mathrm{\overline{MS\kern-0.14em}\kern0.14em}}
\newcommand{\ms}{\mathrm{MS}}

\newcommand{\bm}{b_\mathrm{m}}
\newcommand{\bmu}{b_\mu} 
\newcommand{\bA}{b_\mathrm{A}}
\newcommand{\bp}{b_\mathrm{P}}
\newcommand{\bv}{b_\mathrm{V}}
\newcommand{\bg}{b_\mathrm{g}}

\newcommand{\ca}{c_\mathrm{A}}
\newcommand{\cv}{c_\mathrm{V}}

\newcommand{\csw}{c_\mathrm{csw}}

\newcommand{\za}{Z_\mathrm{A}}
\newcommand{\zp}{Z_\mathrm{P}}
\newcommand{\zv}{Z_\mathrm{V}}
\newcommand{\zm}{Z_\mathrm{m}}

\newcommand{\CF}{C_\mathrm{F}}

\newcommand{\fp}{f_\mathrm{P}}
\newcommand{\fv}{f_\mathrm{V}}

\newcommand{\rz}{\rm I\kern-.2emR}
\newcommand{\gz}{{\rm Z\kern-.35em Z}}
\newcommand{\hz}{\rm I\kern-.2emH}
\newcommand{\pz}{\rm I\kern-.2emP}

\newcommand{\ring}{\mathaccent"7017}

\newcommand{\fm}{\mathrm{fm}}

\newcommand{\gr}{g_{\mathrm{R}}}
\newcommand{\mr}{m_{\mathrm{R}}}
\newcommand{\gbar}{\bar{g}}
\newcommand{\mbar}{\bar{m}}
\newcommand{\Nf}{N_\mathrm{f}}

\newcommand{\Lmax}{L_{\mathrm{max}}}
\newcommand{\FV}{\mathrm{FV}}
\newcommand{\alphaFV}{\alpha_{\FV}}

\renewcommand{\cite}{\shortcite}

\title{Renormalization and lattice artifacts}

\author{P. Weisz}
\affiliation{Max-Planck-Institut f\"{u}r Physik, F\"{o}ringer Ring 6,
D-80805 M\"{u}nchen, Germany}
\authors{1}

\begin{document}

\maketitle

\acknowledgements

I would like to thank Tassos Vladikas and the organizing committee for 
inviting me to present these lectures. I especially thank Rainer Sommer
for his many constructive suggestions for improvement of this manuscript. 
Most of all I would like to thank my wife Teo for all her support 
during my career and for her patience while preparing these lectures. 

\tableofcontents

\maintext


\chapter{Perturbative Renormalization}

\section{Introduction}

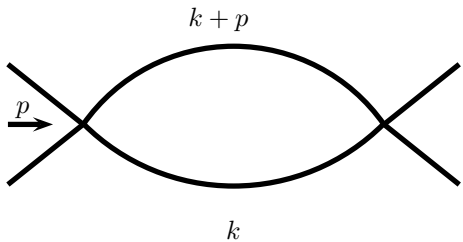
\begin{figure}
\hspace{4.0cm}
\psset{unit=2mm}
\begin{pspicture}(0,-10)(30,10)
\psset{linewidth=2pt}
\psline{->}(0,0)(3,0)
\rput(1,1){$p$}
\qline(0,4)(5,0)
\qline(0,-4)(5,0)
\psarc[showpoints=false](15,-7){12.2}{35}{145}
\psarc[showpoints=false](15,10){14.1}{225}{315}
\rput(14,7){$k+p$}
\rput(15,-7){$k$}
\qline(30,4)(25,0)
\qline(30,-4)(25,0)
\end{pspicture}
\caption{Simple 1--loop diagram} 
\label{oneloop}
\end{figure}

Most of our present knowledge on the structure of renormalization of 
quantum field theories comes from perturbation theory (PT), 
a small coupling expansion around a free field theory. 
In this framework Feynman rules, derived formally in the continuum from 
the Gell-Mann--Low formula 
\shortcite{GellMann:1951rw} or from the path integral 
approach, express amplitudes at a given order as sums of expressions 
associated with Feynman diagrams (see e.g. \shortciteANP{Nakanishi:1971}). 
Tree diagrams are associated with
well defined amplitudes, but for diagrams involving loops 
and associated integration over internal momenta one soon
encounters divergent expressions e.g. for massive $g_0\phi^4$ theory in four 
euclidian dimensions the diagram in Fig.~\ref{oneloop}
is associated with the expression
\be
I=g_0^2\int^\Lambda\rmd^4k\frac{1}{(k^2+m^2)([(k+p)^2+m^2])}
\label{1loop}
\end{equation}  
which is logarithmically divergent as the ultra--violet (UV) cutoff 
$\Lambda\to\infty$. There are many ways to introduce an UV cutoff $\Lambda$
- a specification defines a particular {\it regularization}. 
The process of {\it renormalization} is then a well--defined 
prescription how to map regularized expressions to amplitudes which are 
finite when $\Lambda\to\infty$ while maintaining desired properties
which are summarized by the Osterwalder--Schrader
axioms \footnote{which are equivalent to the Wightman axioms in Minkowski 
space (\shortciteNP{Wightman:1956zz}, \shortciteNP{Streater:1989vi})} 
(\shortciteNP{Osterwalder:1973dx}, \citeyearNP{Osterwalder:1974tc}) 
order by order in PT.

For lattice QCD renormalization is not needed if one is only interested
in obtaining the low energy (LE) spectrum and scattering data. For such 
purposes we need knowledge of the phase diagram, the location of critical 
points where the continuum limit is reached, and the nature of the 
approach e.g. the question whether ratios of masses tend to their 
continuum limit as powers in the lattice spacing $a$:
\be
\frac{m_1(a)}{m_2(a)}=\frac{m_1(0)}{m_2(0)}+C_{12}(am_1)^p\,,\,\,\,p>0\, (?).
\label{artifacts}
\end{equation}
Lattice artifacts will be the topic of chapters~4--6.
But perturbative and non-perturbative renormalization will be needed for
1) computing matrix elements of composite operators (describing probes of 
other interactions, finite temperature transport coefficients,....),
2) relating LE to high energy (HE) scales (e.g. computing running 
couplings and running masses), and 3) giving hints on the nature of 
lattice artifacts.

In this chapter we shall mainly consider perturbative renormalization;
in chapter~2 we shall discuss renormalization group equations which follow
from multiplicative renormalization, and aspects of non-perturbative  
renormalization will be the subject of chapter~3.

\section{History and basic concepts}

In a series of papers Reisz 
(\citeyearNP{Reisz:1987da}--\citeyearNP{Reisz:1988kk})
has proven perturbative renormalizability of 
lattice Yang--Mills theory and QCD (for a large class of actions).
His proof is based mainly on methods developed with continuum 
regularization, (with some important modifications that we will mention 
later), so we will start the discussion with these.

Renormalization theory has a long and interesting history; here I reproduce
a part of Wightman's delightful discussion \shortcite{Wightman:1975gi} 
since the reference is not always easily available. \\
{\it ``One of the first steps in natural history is to establish 
a classifactory nomenclature. I will do this for perturbative renormalization
theory, but in so doing, I want to tell stories with a moral for an earnest 
student. Renormalization theory has a history of egregious errors by 
distinguished savants. It has a justified reputation for perversity; 
a method that works up to 13th order in the perturbation series fails in the
14th order. Arguments that sound plausible often dissolve into mush when 
examined closely. The worst that can happen often happens. The prudent
student would do well {\bf to distinguish sharply between what has been 
proved and what has been made plausible,} and in general he should watch out!}

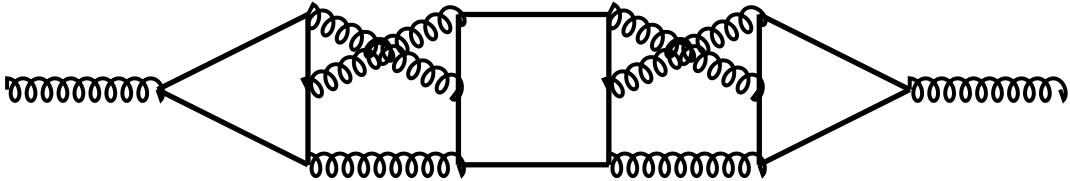
\begin{figure}
\psset{unit=2mm}
\begin{pspicture}(0,-10)(70,10)
\psset{linewidth=2pt}
\pscoil[linewidth=1.5pt,coilarm=0,coilwidth=1.5]{-}(0,0)(10,0)           
\qline(10,0)(20,5)         
\qline(10,0)(20,-5)        
\qline(20,5)(20,-5)
\pscoil[linewidth=1.5pt,coilarm=0,coilwidth=1.5]{-}(20,5)(30,0)
\pscoil[linewidth=1.5pt,coilarm=0,coilwidth=1.5]{-}(20,0)(30,5)
\pscoil[linewidth=1.5pt,coilarm=0,coilwidth=1.5]{-}(20,-5)(30,-5)
\qline(30,5)(30,-5)
\qline(30,5)(40,5)
\qline(30,-5)(40,-5)
\qline(40,5)(40,-5)
\pscoil[linewidth=1.5pt,coilarm=0,coilwidth=1.5]{-}(40,0)(50,5)
\pscoil[linewidth=1.5pt,coilarm=0,coilwidth=1.5]{-}(40,-5)(50,-5)
\pscoil[linewidth=1.5pt,coilarm=0,coilwidth=1.5]{-}(40,5)(50,0)
\qline(50,5)(50,-5)
\qline(50,5)(60,0)
\qline(50,-5)(60,0)
\pscoil[linewidth=1.5pt,coilarm=0,coilwidth=1.5]{-}(60,0)(70,0)
\end{pspicture}
\caption{Troublesome $14$'th oder QED diagram} 
\label{figtrouble}
\end{figure}

{\it My first cautionary tale has to do with the early days of renormalization 
theory. When F.J. Dyson analyzed the renormalization theory of the S--matrix
for quantum electrodynamics of spin one-half particles in his two great 
papers of 1948--9, 
(\shortciteNP{Dyson:1949bp}, \citeyearNP{Dyson:1949ha}) 
he laid the foundations for most later
work on the subject, but his treatment of one phenomenon, overlapping 
divergences was incomplete. Among the methods offered to clarify the 
situation, that of J. Ward \citeyear{Ward:1951} seemed outstandingly simple,
so much so that it was adopted in Jauch and Rohrlich's standard text book
\shortcite{Jauch:1955}. Several years later Mills and Yang noticed that 
unless further
refinements are introduced the method does not work for the photon self
energy \shortcite{Wu:1962zza}. The lowest order for which the trouble 
manifests 
itself is the fourteenth e.g. in the (7--loop) graph Fig.~\ref{figtrouble}.
Mills and Yang repaired the method and sketched some of the steps
in a proof that it would yield a finite renormalized amplitude 
\shortcite{Mills:1966vn}.
An innocent reading of the textbook of Jauch and Rohrlich, 
would never suspect such refinements are necessary.

Another attempt to cope with the overlapping divergences was made by Salam
(\citeyearNP{Salam:1951sm}, \citeyearNP{Salam:1951sj}). 
I will not describe it, if for no other reason than that 
I never have succeeded in understanding it. Salam and Matthews 
commenting on this and related work somewhat later 
\citeyear{Matthews:1951sk} remarked 
``... The difficulty, as in all this work is to find a notation which is both
concise and intelligible to at least two people of whom one may be the 
author". The belief is widespread that when Salam's work is combined
with later significant work by S. Weinberg \citeyear{Weinberg:1959nj}, 
the result should be
a mathematically consistent version of renormalization theory. At least that
is what one reads in the text book of Bjorken and Drell for quantum
electrodynamics \shortcite{Bjorken:1965}, and in the work of R. Johnson 
\citeyear{Johnson:1970it}
and the lectures of K. Symanzik for meson theories 
\shortcite{Symanzik:1961}.
So apparently the Matthews--Salam criterion has been satisfied. I only
wish they had spelled it out a little for the peasants.

Another foundation of renormalization theory with a rather different 
starting point was put forward by Stueckelberg and Green 
\citeyear{Stueckelberg:1951}. 
It was refounded and brought to a certain stage of completion in the
standard text book of Bogoliubov and Shirkov \citeyear{Bogoliubov:1959}.
The mathematical nut that had to be cracked is in the paper of 
Bogoliubov and Parasiuk \citeyear{Bogoliubov:1957gp}, 
(amazingly, not quoted in 
the English translation of Bogoliubov and Shirkov)
\footnote{see also \shortcite{Parasiuk:1960}.}. 
This paper introduces
a systematic combinatorial and analytic scheme for overcoming 
the overlapping divergence problem. This paper is very important
for later developments. Unfortunately it was found by K. Hepp 
\citeyear{Hepp:1966eg}
that Theorem 4 of the paper is false, and that consequently the
proof of the main result is incomplete as it stands. However Hepp found 
that Theorem 4 is not essential to derive the main result and he could fill
all the gaps. Thus it is appropriate to introduce the initials BPH
to stand for the renormalization method described in 
\shortcite{Bogoliubov:1957gp} and \shortcite{Hepp:1966eg}. 
So far as I know it was the first version of renormalization 
theory on a mathematically sound basis."}

The rest of the article is also highly recommendable. Reading his article 
one can understand why many workers in the field who had gone through
these historical developments were wary about renormalization theory
\footnote{including my first lecturer (in 1966) on QFT, 
who began his lectures by saying
``There are only about 3 people in the world who really understand
renormalization theory, and I am not one of them!"}. 
A real breakthrough in the
ease of understanding was supplied by Zimmermann 
(\citeyearNP{Zimmermann:1968mu}, \citeyearNP{Zimmermann:1972te},
 \citeyearNP{Zimmermann:1972tv}),
(in particular also concerning renormalization of composite operators).
I strongly recommend Lowenstein's article \citeyear{Lowenstein:1975ug} 
for a clear exposition of Zimmermann's methods.

It is the purpose of this chapter to give an 
overview of the important steps and concepts of perturbative 
renormalization, stating the main results without proofs; for the latter the 
interested reader must consult further literature.
Moreover I will only discuss the standard approach using expansions in
Feynman diagrams, but I would like to mention a powerful alternative approach 
using flow equations based on the renormalization group developed by Wilson
(\citeyearNP{Wilson:1971bg}, \citeyearNP{Wilson:1971dh}, \citeyearNP{Wilson:1973jj}) 
and improved by Polchinski \citeyear{Polchinski:1983gv}, 
a framework which is
well suited for proving structural results, 
see e.g. the works of Keller, Kopper and Salmhofer 
\citeyear{Keller:1990ej}, \shortcite{Salmhofer:1999uq}.

\section{Perturbative classification of theories}

Concerning the nature of a Feynman diagram 
the first important concept is its {\it superficial degree of divergence}.
Consider a diagram with $E$ external lines, $V$
internal vertices, $P$ internal lines and $\ell$ independent loops.
There are some relations between these the first of which is 
the topological relation
\footnote{A (sub)diagram is connected if all pairs of its vertices are
connected by a path following internal lines.}
\begin{equation}
V+\ell-P=\mathrm{number\,\,of\,\,connected\,\,components}\,.
\end{equation}
Other relations depend on the precise nature of the vertices in the theory  
e.g. with only a 4--point vertex: 
\be
4V=2P+E\,.
\end{equation}
Then for a connected diagram $\Gamma$ with $\ell$ loops the superficial
degree of divergence $\delta(\Gamma)$ estimates the behavior of the 
associated integral when all internal momenta are large. For $\phi^4_d$
theory (in $d$ space--time dimensions):
\ba
\delta(\Gamma)&=&d\ell-2P
\\
&=&d-\frac12(d-2)E+(d-4)V\,.
\ea
For $d=4$ this simplifies to $\delta(\Gamma)=4-E\,,$ which is negative
for $E>4$ indicating possible overall convergence of the integral
in this case.

\begin{exercise}
Show $\delta(\Gamma)=6-2E$ for $\phi^3_6$ theory.
\end{exercise}
This theory is often discussed in the literature because of the 
comparative simplicity of the diagrams involved.
In this case $\delta(\Gamma)$ is negative for $E>3$. 

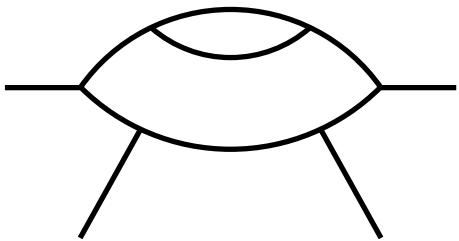
\begin{figure}
\hspace{4.0cm}
\psset{unit=2mm}
\begin{pspicture}(0,-10)(30,10)
\psset{linewidth=2pt}
\qline(0,0)(5,0)
\qline(30,0)(25,0)
\psarc[showpoints=false](15,-7){12.2}{35}{145}
\psarc[showpoints=false](15,10){8}{229}{311}
\psarc[showpoints=false](15,10){14.1}{225}{315}
\qline(5,-10)(9,-2.8)
\qline(25,-10)(21,-2.8)
\end{pspicture}
\caption{2--loop diagram in $\phi^3_6$ theory with a nested divergence} 
\label{twoloop}
\end{figure}

It is clear that considerations of 
$\delta(\Gamma)$ alone are not sufficient to establish 
convergence of the integral because $\Gamma$ could have divergent
sub-diagrams as in Fig.~\ref{twoloop}.  
However it does lead to the following classification of perturbative QFT:

\begin{itemize}

\item {\it super--renormalizable}: theories where only a finite number of
diagrams have $\delta\ge0$.

\item {\it renormalizable}: there exists $E_0<\infty$ such that all 
diagrams with $E>E_0$ have $\delta<0$.

\item {\it non--renormalizable}: all $E$--point diagrams 
have $\delta\ge0$ for sufficiently high number of loops $\ell$.

\end{itemize}

{\it Remarks:} i). Non-renormalizable theories are not necessarily 
unphysical or devoid of predictive power (an interesting example
concerns gravity \shortcite{Donoghue:1995cz}). 
They can be good effective theories, such as chiral 
Langrangians discussed in Goltermann's lectures. 

ii) Non--perturbative formulations of perturbatively renormalizable  
theories may be {\it trivial} in the limit that the ultra--violet
cutoff is sent to infinity
\footnote{the usual perturbative assumption that a finite coupling
exists in the continuum limit is not fulfilled}. Examples of such 
theories are thought to be $\phi^4$ and QED in 4--dimensions,
albeit there is at present no rigorous proof of this conventional wisdom.

iii) There are some rigorous non-perturbative constructions of 
super--renormalizable theories e.g. $\phi^4_2$ \shortcite{Glimm:1968kh}, 
Yukawa theory in 2--dimensions \shortcite{Seiler:1975gs}, 
$\phi^4_3$ \shortcite{Feldman:1975da}.
The Schwinger functions in the topologically trivial sector 
of SU(2) Yang--Mills theory in 4--dimensions in a sufficiently 
small volume have also been rigorously constructed 
(\shortciteNP{Magnen:1992ww}, \citeyearNP{Magnen:1992wv}). 
For all of these cases the structural information on 
renormalization found in perturbation theory carry over to the 
non--perturbative framework.

In the following we restrict attention to renormalizable theories.
A generic procedure to put perturbative renormalization on a firm
mathematical basis consists of three steps:

1. Introduce an UV regularization.

2. Construct a mapping of a Feynman integral to an absolutely convergent
integral when the cutoff is removed.

3. Show that the map in step 2 is equivalent to a renormalization of
the bare parameters and fields in the original Lagrangian, which formally
ensures the desired axiomatic properties of the resulting amplitudes.

There are many perturbative regularization procedures
\footnote{Some involving exotic mathematics such as quantum groups, 
$p$-adic numbers,...}, each with their own
advantages and disadvantages. In all cases known they give equivalent 
physical results.
 
\section{Zimmermann's forest formula}

Let us first restrict attention to theories with massive bare propagators.
Returning to the simple 1--loop example (\ref{1loop}), the integral can 
obviously be made convergent if the integrand is subtracted at 
external momentum $p=0$:
\be
I\rightarrow R=g_0^2\int^\Lambda\rmd^4k
\left[\frac{1}{(k^2+m^2)([(k+p)^2+m^2)}-\frac{1}{(k^2+m^2)^2}\right]\,.
\end{equation}  
We note that since the subtraction is a constant 
(independent of the external momenta)
it can be absorbed in a renormalization of the bare coupling $g_0$. 

The problem now is to find the generalization of the 
subtraction in the simple case above for an arbitrary diagram. 
For this purpose we can restrict attention to {\it one particle 
irreducible (1PI) diagrams}, diagrams which remain connected after cutting
any internal line, since an arbitrary diagram can be constructed from 1PI 
parts joined by propagators.

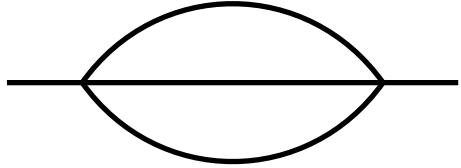
\begin{figure}
\hspace{4.0cm}
\psset{unit=2mm}
\begin{pspicture}(0,-10)(30,10)
\psset{linewidth=2pt}
\qline(0,0)(30,0)
\psbezier{-}(5,0)(10,7)(20,7)(25,0)
\psbezier{-}(5,0)(10,-7)(20,-7)(25,0)
\end{pspicture}
\caption{2--loop diagram with overlapping divergences} 
\label{overlapping}
\end{figure}

Here we consider mappings where subtractions are made directly on the
integrands, as adopted by Reisz for the lattice regularization
\footnote{In the case of dimensional regularization the subtraction
is simpler since the subtractions are in this case $\epsilon$--pole parts
of sub-integrals.}. Let $I_\Gamma(p,k)$ be the original integrand of a 
diagram with external momenta $p=\{p_1,p_2,\dots\}$ 
and internal momenta $k=\{k_1,k_2,\dots\}$.
The regularized integral will then be given by
\be
R_\Gamma(p)=\int\prod_{j}\rmd^d k_j\,R_\Gamma(p,k)\,,
\end{equation}
with
\ba
R_\Gamma(p,k)&=&I_\Gamma(p,k)-\mathrm{subtractions}
\\
&=&\left(1-t_p^{\delta(\Gamma)}\right)\overline{R}_\Gamma(p,k)\,,
\ea
where $t_p^\delta$ is the Taylor operator of degree $\delta$:
\be
t_p^\delta F(p)=F(0)
+p_i^\mu\frac{\partial}{\partial p_i^{\prime\mu}}F(p')|_{p'=0}
+\dots
+\frac{1}{\delta!} p_{i_1}^{\mu_1}\dots p_{i_\delta}^{\mu_\delta}
\frac{\partial}{\partial p_{i_1}^{\prime\mu_1}}\dots
\frac{\partial}{\partial p_{i_\delta}^{\prime\mu_\delta}}F(p')|_{p'=0}\,.
\end{equation}
Dyson's original proposal \citeyear{Dyson:1949ha} was
\be
\overline{R}_\Gamma(p,k)=\prod_{\gamma,1PI\subset\Gamma}
\left(1-t_{p^\gamma}^{\delta(\gamma)}\right)I_\Gamma(p,k)
\nonumber
\end{equation}
where $p^\gamma$ are the external momenta of $\gamma$. On the rhs of this 
equation the term involving $\gamma_1$ is to the right of that for 
$\gamma_2$ if $\gamma_1$ is {\it nested} in $\gamma_2$ 
i.e. $\gamma_1\subset\gamma_2$.
If sets of lines of $\gamma_1$ and $\gamma_2$ do not intersect then 
the order is irrelevant. The proposal cures the problem of nested divergences, 
however no prescription is given how to order {\it overlapping divergences}, 
that is divergent sub-diagrams which have non-trivial 
intersection but are not nested such as in Fig.~\ref{overlapping}
\footnote{It is amusing to hear Dyson's account of his realization
of the problem at http:webstories.com}. 

\begin{figure}
\hspace{4.0cm}
\psset{unit=2mm}
\begin{pspicture}(0,-10)(30,10)
\psset{linewidth=2pt}
\qline(0,0)(5,0)
\qline(30,0)(25,0)
\psarc[showpoints=false](15,-10){14.1}{45}{135}
\psarc[showpoints=false](15,10){14.1}{225}{315}
\qline(12,-3.7)(12,3.7)
\qline(18,-3.7)(18,3.7)
\end{pspicture}
\caption{3--loop diagram in $\phi^3_3$ theory} 
\label{threeloop}
\end{figure}
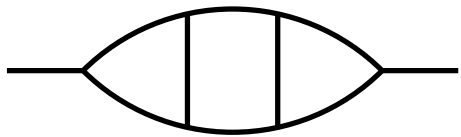

The correct prescription presented by Zimmermann \citeyear{Zimmermann:1968mu}
is obtained by expanding Dyson's product and dropping all terms 
containing products of terms involving overlapping sub-diagrams. 
The result is called {\it Zimmermann's forest formula}:
\be
\overline{R}_\Gamma(p,k)=\sum_{\calF}\prod_{\gamma\in\calF}
\left(-t_{p^\gamma}^{\delta(\gamma)}\right)I_\Gamma(p,k)
\end{equation}
where the sum is over all forests: 
\ba
\calF&=&\{\gamma |\gamma {\rm\,\,1PI\,,\,}\delta(\gamma)\ge0;
\nonumber\\
&&\gamma_1,\gamma_2\in\calF\Rightarrow \gamma_1\subset\gamma_2\,\,
\mathrm{or}\,\,
\gamma_2\subset\gamma_1\,\,\mathrm{or}\,\,\gamma_1,\gamma_2\,\,
\mathrm{non-overlapping}\}\,.
\nonumber
\ea
The empty set $\phi$ is included as a special forest.
\begin{exercise}
Give the set of forests for the diagram in 
Fig.~\ref{threeloop} (it has 8 elements).
\end{exercise}
The following rules are used to specify the momenta flow in $\Gamma$ 
and its subgraphs. The momenta flowing in the $\sigma$'th internal line
$L_{ab\sigma}$ joining vertices $a,b$ is specified as
$l_{ab\sigma}=P_{ab\sigma}(p)+K_{ab\sigma}(k)$ with
$K_{ab\sigma}=\sum_i \epsilon_{ab\sigma i}k_i$;   
$\epsilon_{ab\sigma i}=1$ if $l_{ab\sigma}\in$ loop  ${\calC}_i$;
$\epsilon_{ab\sigma i}=-1$ if $l_{ba\sigma}\in$ loop  ${\calC}_i$;
$\epsilon_{ab\sigma i}=0$ otherwise.
The momentum $P_{ab\sigma}$ is solved by Kirchoff's laws: 
$\sum_{b\sigma} P_{ab\sigma}+p_a=0$. $\sum_{L_{ab\sigma}\in{\calC}_i}
r_{ab\sigma}P_{ab\sigma}=0$. Here ``resistances'' $r_{ab\sigma}$ can be 
chosen for convenience (including zero) but no closed 
loop of zero resistances.

Similarly for subdiagrams $\gamma$: 
$l^\gamma_{ab\sigma}=P^\gamma_{ab\sigma}(p^\gamma)+K^\gamma_{ab\sigma}(k)$. 
First solve for $P^\gamma_{ab\sigma}$ keeping same resistances
and momentum conservation at the vertex. Then determine the
$K^\gamma_{ab\sigma}(k)$ from $l^\gamma_{ab\sigma}=l_{ab\sigma}$.

After applying the forest formula the resulting integrand has a form
\be
F(p,k)=\frac{V(k,p,m)}{\prod_{i=1}^n\left(l_i^2+m^2\right)}\,, 
\end{equation}
where 
\be
l_i=K_i(k)+P_i(p)\,.
\end{equation}
To proceed it is first necessary to give conditions on such an integrand
required to guarantee the convergence of the corresponding integral
\be
F(p)=\int\rmd^d k_1\dots\rmd^d k_\ell\,F(p,k)\,.
\label{Fp}
\end{equation}
Here and in the following we will assume that the internal momenta
have been selected in a ``natural way'' so that $K_i=k_i$ for 
$i=1,\dots,\ell$ (this is always possible). Consider the set
\be
\calL=\{K_1,\dots,K_n\}\,,
\end{equation}
and let $u_1,\dots,u_r,v_1,\dots,v_{\ell-r}\,\,\,r\ge1\,,$ be $\ell$ 
linearly independent elements of $\calL$. Let $H$ be the (Zimmermann) 
subspace spanned by the $u_i$. Then the {\it upper degree} 
of a function $f(u,v)$ with respect to $H$ is given by
\be
\udeg_{u|v} f(u,v)=\overline{\nu}
\,\,\,\,\,\mathrm{if}\,\,\,\,\,
\lim_{\lambda\to\infty}\lambda^{-\overline{\nu}}f(\lambda u,v)\ne0,\infty\,,
\end{equation}
and for the integral (\ref{Fp})
the upper degree wrt $H$ is defined by
\be
\udeg_H F(p)=dr+\udeg_{u|v}F(p,k(u,v,p))\,.
\end{equation}

The important statement is
{\it Dyson's power counting theorem}: $F(p)$ in (\ref{Fp}) is absolutely
convergent if $\udeg_H F(p)<0$ for all subspaces $H$.

The theorem was first proven by Weinberg \citeyear{Weinberg:1959nj}, 
and later a simpler proof was given by Hahn and Zimmermann 
\citeyear{Hahn:1968}.

It can be shown that the forest formula yields integrals 
$F(p)=R_\Gamma(p)$ whose integrands satisfy the conditions of the theorem.
Once this is done it remains to prove step 3 of the renormalization
procedure (see sect.~1.3). This is not at all obvious from the 
forest formula as it stands. For this purpose it is advantageous 
to show that the forest formula is a solution to the 
{\it Bogoliubov--Parasiuk--Hepp recursion relations} 
(\shortciteNP{Bogoliubov:1959}, \shortciteNP{Bogoliubov:1957gp},
 \shortciteNP{Hepp:1966eg}) 
\footnote{It has been noted that the BPH R--operation has the structure 
of a Hopf algebra; see e.g. Kreimer's book \citeyear{Kreimer:2000zh}}:
\be
\overline{R}_\Gamma=\sum_\psi\prod_{\gamma\in\psi}(-t^{\delta(\gamma)})
\overline{R}_\gamma\,,
\end{equation}
where $\psi$ are sets of disjoint 1PI subgraphs.
This formula serves as a basis for an inductive proof of step 3;
again the proof is too long to present here
\footnote{For slightly simplified proofs of BPH renormalization
see \shortcite{Caswell:1981ek}.}; 
a tactic to prove multiplicative renormalizability of gauge 
theories will be given in sect.~1.6.

\section{Renormalization of theories with massless propagators}

For theories with massless bare propagators one must control possible
infra-red (IR) divergences. To this end for a Zimmermann subspace 
one defines a {\it lower degree}:
\be
\ldeg_{v|u}f(u,v)=\underline{\nu}
\,\,\,\,\,\mathrm{if}\,\,\,\,\,
\lim_{\lambda\to0}\lambda^{-\underline{\nu}}f(\lambda u,v)\ne0,\infty\,,
\end{equation}
and for the integral (\ref{Fp}):
\be
\ldeg_{H'} F(p)=d(\ell-r)+\ldeg_{v|u} F(p,k(u,v,p))\,.
\end{equation}
The appropriately modified power counting theorem then states that
$F(p)$ is absolutely convergent if $\udeg_H F(p)<0$ and 
$\ldeg_{H'}F(p)>0$ for all $H$ and for all {\it 
non--exceptional external momenta} $p$ (that is a set for which no partial 
sum of external momenta vanish, except for the complete sum expressing 
total momentum conservation).

To achieve this one has to modify the $R$--operation described above to 
avoid IR divergences introduced by Taylor subtractions at $p=0$.
One way, introduced by Lowenstein and Zimmermann \citeyear{Lowenstein:1975rg} 
and adopted by Reisz (\citeyearNP{Reisz:1987pw}, \citeyearNP{Reisz:1987hx})
is to introduce an auxiliary mass in the massless propagators $l^2\to 
l^2+(1-s)\mu^2$. UV subtractions are then made at $p=0,s=0$, and 
afterwards $s$ is set to 1 in the remaining parts. 
In order to avoid further IR singularities it is usually necessary to
make extra finite subtractions to ensure 2-- and 3--point
functions vanish at exceptional external momenta singularities. 
In gauge theories this is guaranteed by the Slavnov--Taylor identities
which we consider in the next section.
 
\section{Renormalization of 4-d Yang--Mills theory}

In this section we will outline the proof of multiplicative 
renormalizability of Yang--Mills (YM) theory in 4 dimensions.
Perturbation theory is an expansion around a saddle point. For gauge 
theories, such as pure YM theory, gauge invariance of the action 
implies severe degeneracy of the saddle point. To apply perturbation 
theory one needs to lift the degeneracy by {\it gauge fixing} 
(see L\"{u}scher's elegant discussion \citeyear{Luscher:1988sd}) 
\footnote{For gauge invariant correlation functions it is
in principle possible to do PT in finite volumes without gauge fixing
by solving systems of coupled Schwinger--Dyson equations. 
But in practice this is not usually done
because the system is difficult to solve analytically 
(without hints of the structure of the associated Feynman diagrams).}. 
It is convenient to employ linear gauge fixing functions, 
e.g. the Lorentz gauge, which transform under the
adjoint representation of the global gauge group so that the gauge fixed 
action is invariant under such transformations.
Because of the non-linearity of the YM theory this way of fixing 
the gauge introduces extra terms in the functional
integral, which can be expressed as a local Lagrangian involving
{\it Faddeev-Popov} ghost fields $c,\bar{c}$ as introduced in 
Hernandez' lectures. The action in this case is
\ba
S_0&=&-2\int\tr\left\{\frac14 F^2+\frac{\lambda_0}{2}(\partial A)^2
-\bar{c}\partial_\mu \left[D_\mu,c\right]\right\}
\\
&=&S_A+S_\mathrm{gf}+S_\mathrm{FP}\,,
\ea
where $D_\mu=\partial_\mu+g_0 A_\mu$ and 
$F_{\mu\nu}=\partial_\mu A_\nu-\partial_\nu A_\mu
+g_0\left[A_\mu,A_\nu\right]\,.$ Here $A_\mu=T_d A_\mu^d$ where $T_d$ 
denote a basis of the Lie algebra of SU($N$) with 
$\tr(T_dT_e)=-\frac12\delta_{de}$.

The action is no longer gauge invariant but an important property is that
$S_A$ and $S_\mathrm{gf}+S_\mathrm{FP}$
are separately invariant under {\it BRST transformations} 
\shortcite{Becchi:1975nq}:
\ba
\delta A_\mu&=&\epsilon sA_\mu=\epsilon \left[D_\mu,c\right]\,, 
\\
\delta c&=&\epsilon sc =-\epsilon g_0 c^2\,,
\\
\delta\overline{c}&=&\epsilon s\overline{c}=
\epsilon\lambda_0\partial_\mu A_\mu\,,
\ea
where $\epsilon$ is an infinitesimal Grassmann parameter.
\begin{exercise}
Show the BRST invariance of $S_\mathrm{gf}+S_\mathrm{FP}$.  
\end{exercise}

Note that $s$ is nilpotent: $s^2 H[A,\bar{c},c]=0$ for all functionals $H$.
This property and the BRST invariance of the action and measure imply 
the powerful {\it Slavnov--Taylor identities} 
\shortcite{Slavnov:1972fg}, \shortcite{Taylor:1971ff}
\footnote{According to Muta \citeyear{Muta:1987mz}, these generalized WI 
were first 
discussed by 't Hooft \citeyear{'tHooft:1971fh}. In his discussion only
a part of the system of identities was taken into account};
for any functional $H[A,\bar{c},c]$:
\be
\langle s H\rangle=0\,.
\end{equation} 
For example if we take $H=\bar{c}^a(x)\partial_\nu A_\nu^b(y)$ we obtain
\ba
\langle\lambda_0\partial_\mu A_\mu^a(x)\partial_\nu A_\nu^b(y)\rangle
&=&\langle\bar{c}^a(x)\partial_\nu\left[D_\nu,c\right]^b(y)\rangle
\\
&=&\delta^{ab}\delta(x-y)\,,
\ea
where in the last line the ghost field equation of motion has been used.
It follows that the full (bare) gluon propagator has the form
\be
\widetilde{D'}^{ab}_{\mu\nu}(k)=\frac{\delta_{ab}}{k^2}
\left[\frac{\delta_{\mu\nu}-k_\mu k_\nu/k^2}{1+\Pi(k^2)}
+\frac{1}{\lambda_0}\frac{k_\mu k_\nu}{k^2}\right]\,.
\label{fullgprop}
\end{equation}

More generally BRST invariance gives relations between bare Green 
functions and those containing composite operator insertions which 
describe the non-linear symmetry transformations of the fields. 
Many features of the theory derive from these Ward identities. 
In particular they are {\it sufficient to restrict the structure of 
the counter-terms and show multiplicative renormalizability} 
i.e. the theory is renormalized by adjusting 
the bare parameters and fields of the bare action. 
Also the renormalized theory 
satisfies similar symmetry properties as the bare one
\footnote{since the symmetry transformations are 
multiplicatively renormalized,}, which is also crucial to prove the IR 
finiteness of correlation functions with non-exceptional kinematics.

To proceed systematically the WI's are summarized
in equations for generating functionals 
\shortcite{Becchi:1975nq}, \shortcite{ZinnJustin:2002ru} such as 
\ba
Z_0(J,\bar{\xi},\xi,K,L;g_0,\lambda_0)&=&N_0\int D[A]D[c]D[\bar{c}]
\exp\Bigl\{{\calE}_0(A,c,\bar{c},K,L;g_0,\lambda_0)
\nonumber\\
&+&\frac{1}{\hbar}S_c(A,c,\bar{c};J,\bar{\xi},\xi)\Bigr\}\,,
\ea
where $N_0$ is chosen such that 
$Z_0(\mathrm{all\,\,sources}\,=0;g_0,\lambda_0)=1$. 
The term $S_c$ contains source terms for the basic fields
\be
S_c(A,c,\bar{c};J,\bar{\xi},\xi)
=\int_x\sum_b\left\{ J_\mu^b(x)A^b_\mu(x)+\bar{\xi}^b(x)c_b(x)
+\bar{c}^b(x)\xi_b(x)\right\}\,,
\end{equation}
and 
\ba
{\calE}_0(A,c,\bar{c},K,L;g_0,\lambda_0)&=&
\frac{1}{\hbar}\int_x\sum_b\left\{K^b_\mu(x)s A_\mu^b(x)
+L^b(x) s c^b(x)\right\}
\nonumber\\
&&-\frac{1}{\hbar}S_0(A,c,\bar{c};g_0,\lambda_0)\,,
\ea
where the new sources $K,L$ must be introduced for the composite fields 
appearing in the BRS transformations. Note that since $s$ is nilpotent 
further sources in $Z_0$ are not needed. The factors of $\hbar$ in 
diagrammatic contributions to correlation functions count the 
number of associated loops.

Now the BRS transformations imply that $Z_0$ satisfies
\be
\int_x\sum_b\left\{ 
J_\mu^b(x)\frac{\partial}{\partial K_\mu^b(x)}
-\bar{\xi}_b\frac{\partial}{\partial L^b(x)}
-\lambda_0\xi^b(x)\partial_\mu\frac{\partial}{\partial J^b_\mu(x)}
\right\}Z_0=0\,,
\end{equation}
and the same equation holds for the generating functional of the 
connected Green functions $W_0=\hbar\ln Z_0$. For considerations of 
renormalizability we have already noted that it
is more convenient to consider 1PI diagrams. The generating
functional $\Gamma_0$ for these is called the {\it vertex functional} 
(see Hernandez' lectures) which is obtained from $W_0$ via 
a Legendre transformation,
\ba
&&W_0(J,\bar{\xi},\xi,K,L;g_0,\lambda_0)
=\Gamma_0(A,c,\bar{c},K,L;g_0,\lambda_0)
\nonumber\\
&+&\int_x\left\{J^b_\mu(x)A^b_\mu(x)+\bar{\xi}^b(x)c^b(x)
+\bar{c}^b(x)\xi^b(x)\right\}\,,
\ea
where
\be
A_\mu^b(x)=\frac{\partial W_0}{\partial J_\mu^b(x)}\,,\,\,\,
c^b(x)=\frac{\partial W_0}{\partial\bar{\xi}^b(x)}\,,\,\,\,
\bar{c}^b(x)=-\frac{\partial W_0}{\partial \xi^b(x)}\,,\,\,\,
\end{equation}
with inverse relations
\be
J_\mu^b(x)=-\frac{\partial\Gamma_0}{\partial A_\mu^b(x)}\,,\,\,\,
\bar{\xi}^b(x)=\frac{\partial\Gamma_0}{\partial c^b(x)}\,,\,\,\,
\xi^b(x)=-\frac{\partial\Gamma_0}{\partial \bar{c}^b(x)}\,.
\end{equation}
The vertex functional satisfies the equation  
\be
\int_x\left\{\frac{\partial\Gamma_0}{\partial A_\mu^b(x)}
\frac{\partial\Gamma_0}{\partial K_\mu^b(x)}
+\frac{\partial\Gamma_0}{\partial c^b(x)}
\frac{\partial\Gamma_0}{\partial L^b(x)}
-\lambda_0 \partial_\mu A^b_\mu(x))
\frac{\partial\Gamma_0}{\partial \bar{c}^b(x)}
\right\}=0\,.
\label{WI1}
\end{equation}
Setting
\be
\overline{\Gamma}_0=\Gamma_0+\int_x(\lambda_0/2)
\left(\partial_\mu A^b_\mu(x)\right)^2\,,
\end{equation}
and using the FP ghost field equation
\be
\partial_\mu \frac{\partial\overline{\Gamma}_0}{\partial K^b_\mu(x)}
+\frac{\partial\overline{\Gamma}_0}{\partial \bar{c}^b(x)}=0\,,
\end{equation}
we can write the WI in a slightly simpler form
\be
\int_x\left\{\frac{\partial\overline{\Gamma}_0}{\partial A_\mu^b(x)}
\frac{\partial\overline{\Gamma_0}}{\partial K_\mu^b(x)}
+\frac{\partial\overline{\Gamma}_0}{\partial c^b(x)}
 \frac{\partial\overline{\Gamma}_0}{\partial L^b(x)}\right\}=0\,.
\label{FP1}
\end{equation}

We now seek a functional ${\calE}_\rmR={\calE}_0+\rmO(\hbar)$ such that
Green functions generated from  
\ba
Z_\rmR(J,\bar{\xi},\xi,K,L;g_0,\lambda_0)&=&N_\rmR\int D[A]D[c]D[\bar{c}]
\exp\Bigl\{{\calE}_\rmR(A,c,\bar{c},K,L;g,\lambda)
\nonumber\\
&&+(1/\hbar)S_c(A,c,\bar{c};J,\bar{\xi},\xi)\Bigr\}
\ea
are finite order by order in PT
\footnote{Clearly ${\calE}_\rmR$ isn't determined 
uniquely by this condition}. Such a functional can be found and is
simply characterized by the following\\
\noindent {\it Renormalization Theorem:} There exist renormalization 
constants $Z_1, Z_3,\widetilde{Z}_3$ which are formal power series 
in $\hbar$, relating bare fields and renormalized fields
\ba
A^0&=&Z_3^{1/2}A\,,\,\,L^0=Z_3^{1/2}L\,,\,\,\,\lambda_0=Z_3^{-1}\lambda\,,\,\,
g_0=Z_1g\,,
\\
c^0&=&\widetilde{Z}_3^{1/2}c\,,\,\,\bar{c}^0=\widetilde{Z}_3^{1/2}\bar{c}\,,\,\,
K^0=\widetilde{Z}_3^{1/2}K\,,
\ea
such that 
\be
{\calE}_\rmR(A,c,\bar{c},K,L;g_0,\lambda_0)
={\calE}_0(A^0,c^0,\bar{c}^0,K^0,L^0;g_0,\lambda_0)\,.
\end{equation}
Furthermore the $Z$'s have only (multi-) poles in $\epsilon$ with 
dimensional regularization (DR) \shortcite{'tHooft:1972fi},
(\shortciteNP{Breitenlohner:1975hg}, \citeyearNP{Breitenlohner:1976te}),
 \shortcite{Collins:1984xc},
or depend only logarithmically on a momentum cutoff with other regularizations.

The renormalization theorem implies
$\overline{\Gamma}_\rmR(A,c,\bar{c},K,L;g,\lambda)=
\overline{\Gamma}_0(A^0,c^0,\bar{c}^0,K^0,L^0;g_0,\lambda_0)$
and the renormalized vertex functional satisfies the same equation as
$\Gamma_0$. 
The form of the WI indicates universality of the renormalized theory,
(the correlation functions being fixed by 
a finite number of specified normalization conditions).
The measure and renormalized action
$S_\rmR(A,c,\bar{c},g,\lambda)=S_0(A^0,c^0,\bar{c}^0,g_0,\lambda_0)$
are invariant under 
$\delta_\epsilon A_\mu=Z_A^{-1/2}\delta_\epsilon A_\mu^0\,,\dots$
The same holds true for the contributions to ${\calE}_\rmR$ which are linear 
in $K,L$.

The proof of the theorem is by induction on the number of loops. 
Assume the theorem holds to order $n-1$ in the loop expansion 
and hence the functional $\Gamma_\rmR$ satisfies the identities. 
The key point is that the equations for $\Gamma$ are non-linear,
and one obtains simple equations (involving functional derivatives of the
action) for the divergences in the limit of 
large cutoff which occur at order $n$. Since all subdivergences
have been removed, only overall divergences remain which by
the power counting theorem are polynomials in momentum space of 
the order of the divergence degrees of the corresponding diagrams
\footnote{Degrees of divergence are defined as before.
The superficial degree of divergence of a diagram with 
$N_G,N_{FP},N_F$ external gluon, ghost, and fermion lines is:
$d=4-N_G-\frac32(N_{FP}+N_F)$ have 8 cases with $d\ge0$: 
$(N_G,N_{FP},N_F)=(0,0,0),(2,0,0),(0,2,0),(0,0,2),(3,0,0)
,(1,2,0),(1,0,2),(4,0,0)$. The vacuum diagram can be dropped because it
is absorbed in the normalization of the generating functional.
The rest have degrees $2,1,1,1,1,0,0,0$ respectively. Although superficial
degrees $>0$ the full amplitudes are all only log divergent
due to the gauge symmetry and Lorentz invariance.}.
The equations from the WI suffice to restrict all the coefficients 
(i.e. the renormalization constants) such that the theorem can be 
proven at order $n$. 

From the renormalized BRS WI for the functionals 
follow various relations for correlation functions.
Starting from the identities (\ref{WI1},\ref{FP1}) for $\Gamma_0$ or 
$\Gamma_\rmR$ and differentiations wrt the FP fields and the gauge fields.
As mentioned before one gets e.g. {\it Slavnov-Taylor identities} for 
2-,3-point vertices, which imply absence of mass terms in the gluon 
self--energy and small momenta behavior of 3-point vertex function.

\section{Perturbative renormalization of lattice gauge theory}

For an introduction to perturbation theory with the lattice regularization
please see the lectures of Hernandez. Here I just summarize some of the
salient points. First the lattice action is chosen to be gauge invariant
and such that it has the desired classical continuum limit. The Feynman
rules are algebraically more complicated than in the continuum e.g.
one also encounters vertices involving more than 4 gauge fields, but they are
straightforwardly derived and now usually computer generated. There are also
extra terms coming from the measure (the Jacobian of the change of variables
$U$ to $A$). The integral over the gauge field $A$ is extended to $\infty$,
which formally modifies the integral by only comparably negligible 
non-perturbative terms. For performing the perturbative expansion in this
way, gauge fixing is needed as in the continuum. Finally there is an exact
BRS symmetry on the lattice since this symmetry doesn't originate from 
the continuum aspects but is a general consequence of gauge fixing in the
presence of a gauge invariant cutoff 
(see L\"{u}scher \citeyear{Luscher:1988sd}).

So we have practically all the ingredients to prove 
perturbative renormalizability following the continuum procedure apart from
{\it a power counting theorem for the lattice regularization}.
The latter has been provided by Reisz \citeyear{Reisz:1987da}; very nice 
accounts can be found in \shortcite{Luscher:1988sd} and 
\shortcite{Reisz:1988zv}.
An influential earlier paper considering renormalization of 
lattice gauge theories is by Sharatchandra \citeyear{Sharatchandra:1976af}. 
The difficulty comes from the fact that in momentum space the domain of
integration is compact, internal momenta $k$ are restricted to 
the Brillouin zone $\mathcal{B}$, $|k_\mu|\le\pi/a$, 
and the integrand is a periodic function wrt the loop
momenta. Also, as mentioned above, often the vertices are quite complex and 
there are additional ``irrelevant'' vertices 
i.e. those which vanish in the naive continuum limit. 
In this situation the usual convergence theorems 
in the continuum do not apply.

Let us first consider the case with massive propagators. A general lattice
Feynman diagram corresponds to an integral of the form
\footnote{where for simplicity of notation we consider the case of just one 
mass parameter $m$.}
\be
I=\int_{\mathcal{B}}\rmd^4k_1\dots\rmd^4k_\ell\,\frac{V(k,p,m,a)}{C(k,p,m,a)}\,,
\end{equation}
where the numerator $V$ of the integrand contains all the vertices and
numerators of propagators, and the denominator $C$ is the product
of the denominators of the propagators,

In order to make rigorous statements on the convergence Reisz specified
the following restrictions on $V,C$ which however hold for a large class 
of lattice actions:

\noindent For\,\,$V$:\,\,\,$1)\,\,\,\exists\,\,\,\omega\,\,{\rm st}\,\, 
V(k,p,m,a)=a^{-\omega}F(ak,ap,am)\,,$

\hspace{1.2cm} where $F$ is $2\pi$ periodic in $ak_i$ and a polynomial in $am$.

\hspace{0.6cm} $2)\,\,\lim_{a\to0}V(k,p,m,a)=P(k,p,m)$ exists.

\vspace{0.5cm}
\noindent For\,\,$C$:\,\,\,$1)\,\,\,\,C=\prod_{i=1}^n C_i(l_i,m,a)\,,$ with
\ba
C_i(l_i,m,a)&=&a^{-2}G_i(al_i,am)\,,
\\
l_i(k,p)&=&\sum_j a_{ij}k_j+\sum_l b_{il}p_l\,,\,\,\,a_{ij}\in\gz\,,
\ea
\hspace{1.6cm} the latter (integer) condition maintaining 
the $2\pi$-periodicity of $G_i$.

\hspace{0.6cm} $2)\,\,\,\,\lim_{a\to0}C_i(l_i,m,a)=l_i^2+m^2\,.$

\hspace{0.6cm} $3)\,\,\,\,\exists\,\,\,a_0,A\,\,{\rm st}\,\,|C_i(l_i,m,a)|
\ge A(\hat{l}_i^2+m^2)\,\,\,\,\forall\,\, a\le a_0, \forall\,\,i\,,$

\hspace{1.2cm} the latter condition ensuring that 
$1/|C_i|\sim\rmO(a^2)$ at the boundary

\hspace{1.2cm} of $\mathcal{B}$. 

With these specifications, for $a>0$ and $m^2>0$ the Feynman integrals 
are absolutely convergent and the dependence on external momenta is smooth.
However the lattice integral does not necessarily converge in the 
continuum limit if the continuum limit of the integrand is absolutely
integrable e.g.
\be
\int_{\mathcal{B}}\frac{\rmd^4k}{(2\pi)^4}\,\frac{1-\cos(ak_\mu)}{\hat{k}^2+m^2}
=\frac{1}{8a^2}+\rmO(a^0)\,,
\end{equation} 
i.e. the integral is quadratically divergent although the continuum limit
of the integrand vanishes. 

A lattice degree of divergence involves the behavior 
of the integrand for large internal momenta $k_i$ and simultaneously  
for small $a$. For a Zimmermann subspace (defined as before) the
(upper) degree of divergence is defined by
\ba
\udeg_{u|v}F&=&\bar{\nu}\,\,\,{\rm if}\,\,\,
F(\lambda u,v,m,a/\lambda)\sim_{\lambda\to\infty} k\lambda^{\bar{\nu}}\,,
\\
\udeg_{u|v}I&=&4r+\udeg_{u|v}V-\udeg_{U|V}C\,.
\ea

With this definition of the degree of divergence Reisz proved the
{\it Theorem}: If $\udeg_H I<0$ for all Zimmermann
subspaces, then the continuum limit of $I$ exists and is given by the
corresponding continuum expression.

To prove the renormalizability one must also specify a lattice $R$-operation
corresponding to subtractions of {\it local} lattice counter-terms e.g.
\be
t^\delta\to\hat{t}^\delta\,,\,\,\,
\hat{t}^\delta_p f(p)=f(0)+\ring{p}_\mu\frac{\partial}{\partial p'_\mu}
f(p')|_{p'=0}+\dots
\end{equation}
where $\ring{k}_\mu=a^{-1}\sin(ak_\mu)$ is $2\pi$ periodic in $ak$.

The proof of the theorem is lengthy and cannot be presented here.
Let me just mention that  
the subtractions are organized by the Zimmermann forest formula as
in the continuum, and zero mass propagators are treated as by
Lowenstein and Zimmermann mentioned in sect.~1.5. 
Some fine points involve the treatment of the measure terms. 
The use of the BRS Ward identities also proceeds
similarly to that for continuum regularizations \shortcite{Reisz:1988kk}.

Finally matter fields are included straightforwardly.
Note that criterion 3) is satisfied for Wilson fermions
(which require  an additional additive mass renormalization)
for which $C_i(p)=(1+am)\hat{p}^2+m^2
+\frac12 a^2\sum_{\mu<\nu}\hat{p}_\mu^2\hat{p}_\nu^2$.
It is not satisfied by staggered fermions, however an extension
of the power--counting theorem in this case has been supplied by Giedt
\citeyear{Giedt:2006ib}. This paves the way for a proof of perturbative 
renormalizability of staggered fermions
\footnote{This remains to be done, but Giedt anticipates no additional 
principle difficulties (private communication).}

\chapter{Renormalization group equations}

In the last lecture we learned that correlation functions of the basic 
QCD fields are multiplicatively renormalizable provided bare parameters
are also multiplicatively renormalized. For perturbation theory in the
continuum we usually employ dimensional regularization 
where Feynman rules are developed in $D=4-2\epsilon$ dimensions 
\shortcite{Collins:1984xc}. 
Bare amplitudes have (multi-) poles at $\epsilon=0$ and in the 
{\it minimal subtraction (MS) renormalization scheme} one subtracts just these. 
Renormalization constants then have the form
\be
Z=1+g^2\frac{z_1}{\epsilon}
+g^4\left(\frac{z_2}{\epsilon^2}+\frac{z_3}{\epsilon}\right)+\dots
\end{equation}
where $g$ is the (dimensionless) renormalized coupling related to the bare
coupling by
\be
g^2=\mu^{-2\epsilon}g_0^2Z_g\,,
\end{equation}
with $\mu$ the renormalization scale. Note the MS scheme is a
{\it mass independent renormalization scheme} i.e. a scheme for
which the renormalization constants do not depend on the quark masses. 

Renormalized correlation functions involving $r$ gauge fields, 
$n$ quark-antiquark pairs and $l$ ghost -antighost pairs are given by 
\be
G_\rmR^{r,n,l}(\mu,p;g,\lambda,m_j)=\left(Z_3^{-1/2}\right)^r
\left(Z_2^{-1/2}\right)^{2n}\left(\widetilde{Z}_3^{-1/2}\right)^{2l}
G_0^{r,n,l}(p;g_0,\lambda_0,m_{0j})
\end{equation}
where the renormalized gauge parameter and renormalized quark masses
are given by
\be
\lambda=Z_3\lambda_0\,,\,\,\,\,\,m_j=\zm m_{0j}\,.
\end{equation}

The {\it renormalization group equations} \shortcite{Callan:1970yg}, 
(\shortciteNP{Symanzik:1970rt}, \citeyearNP{Symanzik:1971}), 
follow immediately from the simple observation that bare 
correlation functions are independent of the renormalization scale $\mu$:
\be
\left[\mu\frac{\partial}{\partial\mu}
+\beta\frac{\partial}{\partial g}
+\tau m_j\frac{\partial}{\partial m_j}
+\delta\frac{\partial}{\partial\lambda}
+r\gamma_3+2n\gamma_2+2l\widetilde{\gamma}_3\right]G_\rmR^{r,n,l}=0\,,
\end{equation}
where the coefficient functions are defined through:
\ba
\widetilde{\beta}(\epsilon,g)&=&\mu\frac{\partial g}{\partial\mu}
\vert_{g_0,\lambda_0,m_{0j}}=
-\epsilon g\left\{1-\frac12 g\frac{\partial}{\partial g}\ln Z_g\right\}^{-1}
\\
&=&-\epsilon g+\beta(g)\,,
\\
m_i\tau(g)&=&\mu\frac{\partial m_i}{\partial\mu}\vert_{g_0,\lambda_0,m_{0j}}\,,
\\
\delta(g)&=&\mu\frac{\partial\lambda}{\partial\mu}\vert_{g_0,\lambda_0,m_{0j}}\,,
\\
\gamma_3(g)&=&\frac12\mu\frac{\partial}{\partial\mu}
\ln Z_3\vert_{g_0,\lambda_0,m_{0j}}\,,
\ea
and similarly for $\gamma_2,\widetilde{\gamma}_3$. Note that for any 
function $f$ depending only on $g,\epsilon$,
\be
\mu\frac{\partial}{\partial\mu}f(g,\epsilon)\vert_{g_0,\lambda_0,m_{0j}}
=\widetilde{\beta}(\epsilon,g)\frac{\partial}{\partial g}f(g,\epsilon)\,.
\end{equation}
The functions $\beta$ and $\tau$ have the following perturbative expansions: 
\ba
\beta(g)&=&-g^3\sum_{k=0}^\infty b_k g^{2k}\,,
\\
\tau(g)&=&-g^2\sum_{k=0}^\infty d_k g^{2k}\,,
\ea
with leading coefficients given by
\ba
b_0&=&(4\pi)^{-2}\left[\frac{11}{3}N-\frac23\Nf\right]\,,
\\
b_1&=&(4\pi)^{-4}\left[\frac{34}{3}N^2
-\left(\frac{13}{3}N-\frac{1}{N}\right)\Nf\right]\,,
\\
d_0&=&(4\pi)^{-2}\frac{3(N^2-1)}{N}\,.
\ea
Note that in the MS scheme
$\beta$ and $\tau$ are independent of the gauge parameter $\lambda$.
To show this start from the general relations between renormalized
and bare quantities in the form: 
\ba
g_0&=&\mu^\epsilon\left\{g
+\sum_{r=1}^\infty a_r(g,m_k/\mu,\lambda)\epsilon^{-r}\right\}\,,
\label{g0exp}
\\
m_{0j}&=&m_j
+\mu\sum_{r=1}^\infty b_{jr}(g,m_k/\mu,\lambda)\epsilon^{-r}\,,
\label{m0exp}
\\
\lambda_0&=&\lambda
+\sum_{r=1}^\infty c_r(g,m_k/\mu,\lambda)\epsilon^{-r}\,,
\ea
where the coefficients $a_r,b_{jr},c_r$ are independent of $\epsilon$.
Now define 
\be
\rho\equiv\frac{\partial g}{\partial\lambda_0}/
\frac{\partial\lambda}{\partial\lambda_0}\,,\,\,\,\,\, 
\nu_j\equiv\frac{\partial m_j}{\partial\lambda_0}/
\frac{\partial\lambda}{\partial\lambda_0}\,, 
\end{equation}
where the differentiation wrt $\lambda_0$ is at fixed $g_0,m_{0j},\mu$.
For any renormalized Green function $G$ of gauge invariant operators we
must have an equation of the form:
\be
\left\{\frac{\partial}{\partial\lambda}+\rho\frac{\partial}{\partial g}
+\sum_j\nu_j\frac{\partial}{\partial m_j}+\sigma_G\right\}G=0\,,
\end{equation}
i.e. a change in the gauge parameter must be compensated by a change of the
renormalized parameters and a multiplicative renormalization of the
operators appearing in the definition of $G$. 
Now differentiate eqs.~(\ref{g0exp}), (\ref{m0exp}) wrt $\lambda_0$
to obtain 
\begin{equation}
A
\left(\begin{array}{c}
\rho\\
\mu^{-1}\nu_j\\ 
\end{array}
\right)
+v=0\,,\,\,\,\,
v=\left(\begin{array}{c}
\sum_{r=1}\frac{\partial a_r}{\partial\lambda}\epsilon^{-r}\\
\sum_{r=1}\frac{\partial b_{jr}}{\partial\lambda}\epsilon^{-r}\\
\end{array}
\right)\,,
\end{equation}
where
\begin{equation}
A=
\left(\begin{array}{cc}
1+\sum_{r=1}\frac{\partial a_r}{\partial g}\epsilon^{-r}&
\sum_{r=1}\frac{\partial a_r}{\partial (m_j/\mu)}\epsilon^{-r}\\
\sum_{r=1}\frac{\partial b_{jr}}{\partial g}\epsilon^{-r}&
\delta_{jk}+\sum_{r=1}\frac{\partial b_{jr}}{\partial (m_k/\mu)}\epsilon^{-r}\\
\end{array}
\right)\,.
\end{equation}
Since $\rho$ and $\nu_j$ must be finite we conclude that $A^{-1}v$
must also be finite. It follows that since $a_r,b_{jr}$ do not depend 
on $\epsilon$ we must have
\be
\rho=0\,,\,\,\,\nu_j=0\,\,\forall\,\,j\,,
\end{equation}
and hence
\be
\frac{\partial a_r}{\partial\lambda}=0\,,\,\,\,\,\,
\frac{\partial b_{jr}}{\partial\lambda}=0\,\,\forall\,\,j\,.
\end{equation}
The functions $\beta,\tau_j$ are determined by $a_1,b_{j1}$ and 
are hence also independent of $\lambda$. 

\section{Physical quantities, $\Lambda$--parameter and RGI masses}

A physical quantity $P$ is independent of wave function renormalization
and independent of $\lambda$ and hence satisfies the simplified RG equation:
\be
\left[\mu\frac{\partial}{\partial\mu}
+\beta\frac{\partial}{\partial g}
+\tau m_j\frac{\partial}{\partial m_j}\right]P=0\,.
\label{RGphys}
\end{equation}
Now every solution of this equation can be expressed in terms of special 
solutions. Firstly the $\Lambda$-{\it parameter} which doesn't involve quark
masses: 
\be
\Lambda=\mu\ell(g)\,,
\end{equation}
where $\ell(g)$ satisfies the equation
\be
\left[1+\beta\frac{\partial}{\partial g}\right]\ell(g)=0\,,
\end{equation}
and is completely fixed by its behavior for small $g$:
\be
\ell(g)=(b_0g^2)^{-b_1/(2b_0^2)}\rme^{-1/(2b_0g^2)}
\exp\left\{-\int_0^g\rmd x\,\left[\frac{1}{\beta(x)}+\frac{1}{b_0x^3}
-\frac{b_1}{b_0^2x}\right]\right\}\,.
\end{equation}
The other parameters are the {\it RG invariant masses}
\be
M_i=m_i\theta(g)
\end{equation}
where $\theta(g)$ satisfies
\be
\left[\beta\frac{\partial}{\partial g}+\tau\right]\theta(g)=0\,,
\end{equation}
which is also fixed (i.e. also its normalization) by its behavior as $g\to0$:
\be
\theta(g)=(2b_0g^2)^{-d_0/(2b_0)}
\exp\left\{-\int_0^g\rmd x\,\left[\frac{\tau(x)}{\beta(x)}-\frac{d_0}{b_0x}
\right]\right\}\,.
\end{equation}

Consider as an example the case when $P(Q^2,\mu,g,m_j)$ is a dimensionless
physical quantity depending on a Euclidean momentum $Q$ then:
\be
P=\widetilde{P}\left(Q^2/\mu^2,g,m_j/\sqrt{Q^2}\right)\,.
\end{equation}
We can now appreciate the power of the RG equation; it enables us to 
deduce the behavior of $P$ for large $Q^2$, since it implies
\be
P=\widetilde{P}\left(1,\gbar(t),\mbar_j(t)/\sqrt{Q^2}\right)\,,\,\,\,\,
t\equiv\ln(Q^2/\Lambda^2)\,.
\end{equation}
where $\gbar(t)$ is a {\it running coupling} and $\mbar_j(t)$ is the 
{\it running mass}. The running coupling satisfies the equation
\be
Q\frac{\partial\gbar}{\partial Q}=\beta(\gbar)\,,
\end{equation}
and is implicitly defined by
\be
t=\frac{1}{b_0\gbar^2}+\frac{b_1}{b_0^2}\ln(b_0\gbar^2)+2\int_0^{\gbar}
\rmd x\,\left[\frac{1}{\beta(x)}+\frac{1}{b_0x^3}-\frac{b_1}{b_0^2x}\right]\,.
\end{equation}
As $Q^2\to\infty$ the running coupling tends to zero (logarithmically)
\be
\gbar^2(t)=\frac{1}{b_0 t}\left\{1-\frac{b_1}{b_0^2 t}\ln(t)
+\rmO(t^{-2})\right\}
\end{equation}
a property known as {\it asymptotic freedom}
\shortcite{Gross:1973id}, \shortcite{Politzer:1973fx}.
The running mass is given by
\be
\mbar_j(t)=M_j\theta^{-1}\left(\gbar(t)\right)\,,
\end{equation}
and satisfies the equation
\be
Q\frac{\partial\mbar_j}{\partial Q}=\mbar_j\tau(\gbar)\,.
\end{equation}
For $Q^2\to\infty$ it decreases according to
\be
\mbar_j(t)=M_j\left(\frac{2}{t}\right)^{d_0/(2b_0)}\left\{1
-\frac{d_0b_1}{2b_0^3 t}(1+\ln(t))+\frac{d_1}{2b_0^2t}
+\rmO(t^{-2})\right\}\,.
\end{equation}
Any two mass independent schemes can be related by finite parameter
renormalizations;
\ba
g'&=&g\sqrt{\chi_g(g)}\,,\,\,\,\,\,\chi_g(g)=1+\chi_g^{(1)}g^2+\dots\,,
\\
m'&=&m\chi_{\mathrm{m}}(g)\,.
\ea
Physical quantities are scheme independent and so
(for the case where the renormalization scales are equal)
\be
P(\mu,g,m_j)=P'(\mu,g',m'_j)\,,
\end{equation}
satisfies (\ref{RGphys}) in the first scheme and a corresponding equation in
the second scheme with new coefficients $\beta',\tau'$ which are related to
$\beta,\tau$ by:
\ba
\beta'(g')&=&\left\{\beta(g)\frac{\partial g'}{\partial g}\right\}|_{g=g(g')}\,,
\\
\tau'(g')&=&\left\{\tau(g)+\beta(g)\frac{\partial}{\partial g}
\ln\chi_{\mathrm{m}}(g)
\right\}|_{g=g(g')}\,.
\ea
\begin{exercise}
Show that
it follows that $b'_0=b_0,b'_1=b_1,d'_0=d_0$ i.e. these coefficients are
universal but the higher loop coefficients are not e.g.
\be
d'_1=d_1+2b_0\chi_{\mathrm{m}}^{(1)}-d_0\chi_g^{(1)}\,.
\end{equation}
Also 
\ba
\Lambda'&=&\Lambda\exp\left\{\frac{\chi_g^{(1)}}{2b_0}\right\}\,,
\\
M'&=&M\,,
\ea
i.e. $\Lambda$--parameters are scheme dependent (albeit their relations
require just 1-loop computations), but the {\it RG invariant masses are
scheme independent}.
\end{exercise}

\section{The MS lattice coupling}

With the lattice regularization of QCD we can also define a MS scheme by just 
subtracting powers of logarithms in the lattice cutoff $a$. 
The MS perturbatively renormalized lattice coupling then has the form
\be
g_{\rm latt}=g_0-b_0g_0^3\ln(a\mu)+\dots
\end{equation}
The 2-loop relation between $g_{\rm latt}$ and $g_{\msbar}$ has
been computed for a large class of actions, also including fermions
using sophisticated algebraic programs for automatic generation of the Feynman
rules (see e.g. \shortcite{Luscher:1995np}, \shortcite{Hart:2009nr}, 
\shortcite{Constantinou:2007gv}).
As discussed in the last section, the 1-loop relation gives the ratio 
of $\Lambda$--parameters.
This was first computed for the standard Wilson action by Hasenfratz 
and Hasenfratz \citeyear{Hasenfratz:1980kn};
in that case for $N=3,\Nf=0$ one finds a large number 
$\Lambda_{\msbar}/\Lambda_{\rm latt}=28.8$. Close to the continuum 
limit, assuming that $g_0=0$ is the critical point which should be
approached when taking the continuum limit, 
we should find for a physical mass $m$ in the limit of zero mass quarks
\ba 
&&m/\Lambda_{\rm latt}=c_m+\rmO(a^p)\,,\,\,\,\,\,\,\,
\\ 
&&\Rightarrow ma\sim_{g_0\to0} c(b_0g_0^2)^{-b_1/2b_0^2}
\rme^{-1/(2b_0g_0^2)}R(g_0)\,,\,\,\,\,R(g_0)=1+\rmO(g_0^2)\,.\,\,\,\,\,\,\,
\ea
An observation of the leading behavior above is called ``asymptotic scaling"; 
present spectral measurements are considered consistent with these 
expectations although the correction factor $R(g_0)$ is 
a power series in $g_0$ and hence usually not 
slowly varying in the regions where the simulations are performed.

We remark here again that for lattice regularizations which break chiral
symmetry, such as Wilson fermions (see Hernandez' lectures), 
the quark mass needs also an additive renormalization
\be
m=\zm \mq,\,\,\,\,\mq=m_0-\mc\,.
\end{equation} 
Computation of the quark self energy to 1-loop gives $a\mc,Z_2,\zm$
to $\rmO(g_0^2)$, and subsequent comparison to the result in the $\msbar$
scheme yields $\chi_{\mathrm{m}}^{(1)}$.

\section{Renormalization of composite operators}

Consider a perturbative computation of a correlation function involving
a bare composite operator and a product of basic fields at physically 
separated space-time points. After performing the required renormalization
of the bare parameters and basic fields the resulting expression either 
1) stays finite as the UV cutoff is removed or 2) diverges. If case 1) holds
to all orders of PT then this is (usually) due to the fact that the composite
operator is a conserved current or one satisfying a current algebra. 
In case 2) the simplest situation is that the correlation function 
becomes finite when the composite operator is multiplicatively renormalized,
$$\phi_\rmR(\mu)=Z_\phi(\mu)\phi^{\rm bare}\,.$$ 
But in general operators mix 
with other operators having the same conserved quantum numbers and the same
canonical dimension (or less)
\be
\calO_{\rmR\sigma}=\sum_\tau Z_{\sigma\tau}\calO_\tau^{\rm bare}
+``Z\times{\rm\,\,lower\,\,dimension\,\,ops"}\,.
\end{equation}
A correlation function involving an insertion of a purely multiplicatively
renormalizable operator with a product of diagonally multiplicatively 
(gauge invariant) renormalized operators, all located at physically separated
points, satisfies the RG equation (for the case of massless quarks):
\be
\left[\left(\mu\frac{\partial}{\partial\mu}
+\beta\frac{\partial}{\partial g}
-\sum_i\gamma_{\phi_i}\right)\delta_{\tau\sigma}-\gamma_{\tau\sigma}
\right]G^\sigma_{\rmR;1,\dots,n}=0\,,
\end{equation}
where
\be
\gamma_{\phi_i}=\mu\frac{\partial}{\partial\mu}\ln Z_{\phi_i}\,,\,\,\,\,\,\,
\gamma_{\tau\sigma}=\mu\frac{\partial Z_{\tau\rho}}{\partial\mu}
\left(Z^{-1}\right)_{\rho\sigma}\,.
\end{equation}
For the physical interpretation it is often advantageous to define 
{\it RG invariant operators} by (in the simpler cases)
\be
\phi_{{\rm RGI}\,i}=C_i(\mu/\Lambda)\phi_{\rmR\,i}\,,
\label{RGIop}
\end{equation}
where $C_i$ is a solution to the equation
\be
\left[\mu\frac{\partial}{\partial\mu}
+\beta\frac{\partial}{\partial g}+\gamma_{\phi_i}\right]C_i=0\,.
\end{equation}
It is given (with conventional normalization) by
\be
C_i(\mu/\Lambda)=\left(2b_0\gbar^2(\mu)\right)^{\gamma_{\phi_i}^{(0)}/(2b_0)}
\exp\left\{-\int_0^{\gbar}\rmd x\,\left[\frac{\gamma_{\phi_i}(x)}{\beta(x)}
+\frac{\gamma_{\phi_i}^{(0)}}{b_0 x}\right]\right\}
\end{equation}
where we have assumed that 
$\gamma_{\phi_i}(g)=\gamma_{\phi_i}^{(0)}g^2+\rmO(g^4)$.

As an example of a lattice regularization, let us consider the case
of Wilson's fermions. It has an isovector current 
\ba
V_\mu^a(x)&=&\frac12\Bigl\{
\psibar(x)\frac{\tau^a}{2}(\gamma_\mu-1)U(x,\mu)\psi(x+a\hat{\mu})
\nonumber\\
&&+\psibar(x+a\hat{\mu})\frac{\tau^a}{2}
(\gamma_\mu+1)U(x,\mu)^\dagger\psi(x)\Bigr\}\,,
\ea
where $\tau^a$ are the Pauli matrices acting on the flavor indices,
which is exactly conserved (in the case of degenerate quark masses)
i.e. on shell, $\partial_\mu^* V_\mu^a(x)=0$ where $\partial_\mu^*$
is the lattice backward derivative. It follows that this bare operator 
doesn't need any renormalization. 
No analogous conserved axial vector current exists for Wilson
fermions even for $\mq=0$. Often in practical numerical computations
simpler lattice currents are employed:
\be
V_\mu^a(x)=\psibar(x)\frac{\tau^a}{2}\gamma_\mu\psi(x)\,,\,\,\,\,\,
A_\mu^a(x)=\psibar(x)\frac{\tau^a}{2}\gamma_\mu\gamma_5\psi(x)\,,
\end{equation}
which are expected to be conserved up to lattice artifacts.
In PT this is indeed the case, but in order that the currents 
obey the correct current algebra (up to $\rmO(a)$ artifacts) they 
require a finite renormalization 
\be
V_\rmR=\zv V\,,\,\,\,\,A_\rmR=\za A\,,\,\,\,\,Z_{\rm V}/Z_{\rm A}
=1+\rmO(g_0^2)\,.
\end{equation}

{\it Digression:} It is probably not so well known among students
specializing in lattice theory that there is a problem with ``naive
$\gamma_5$'' in dimensional regularization. Namely the algebraic rules
$$
\left\{\gamma_\mu,\gamma_\nu\right\}=2\delta_{\mu\nu}\,,
\,\,\,\,\,\delta_{\mu\mu}=D\,,
\\
\left\{\gamma_\mu,\gamma_5\right\}=0\,,\,\,\,\,\,\tr\gamma_5=0\,,\,\,\,
{\rm and\,\,cyclicity\,\,of\,\,trace}
$$
imply the unwanted relation
\footnote{To prove this start we use the rules to obtain
$D\tr\gamma_5\gamma_\mu\gamma_\nu=\tr\gamma_5\gamma_\rho^2\gamma_\mu\gamma_\nu
=-\tr\gamma_\rho\gamma_5\gamma_\rho\gamma_\mu\gamma_\nu
=-\tr\gamma_5\gamma_\rho\gamma_\mu\gamma_\nu\gamma_\rho
=(-D+4)\tr\gamma_5\gamma_\mu\gamma_\nu$ so that 
$\tr\gamma_5\gamma_\mu\gamma_\nu=0$ unless $D=2$. Next repeat the manipulations
with the trace involving $\gamma_5$ and 4 gamma matrices.}
$\tr\left(\gamma_5\gamma_\mu\gamma_\nu\gamma_\rho\gamma_\lambda\right)=0$
unless $D=2$ or $D=4$. The modified algebra proposed by 't Hooft and Veltman
\citeyear{'tHooft:1972fi} is to define 
$\gamma_5=\gamma_0\gamma_1\gamma_2\gamma_3$ so that
$$
\left\{\gamma_\mu,\gamma_5\right\}=0\,\,\,{\rm for}\,\,\,\mu\le3\,,\,\,\,
\left[\gamma_\mu,\gamma_5\right]=0\,\,\,{\rm for}\,\,\,\mu>3\,.
$$
The algebra is then consistent but results in the necessity of having
to introduce an infinite renormalization for the bare axial current
\ba
(A_\mu^a)^{\ms}&=&\za^{\ms}(g)\psibar\frac{\tau^a}{2}
\frac12\left[\gamma_\mu,\gamma_5\right]\psi\,,
\\
\za^{\ms}(g)&=&1+g^4\frac{1}{(4\pi)^2} 2b_0\CF\frac{1}{\epsilon}+\dots\,,\,\,\,\,
\CF=\frac{N^2-1}{2N}\,,
\ea
and we need a further renormalization $\chi_A(g)=1-g^2\frac{4}{(4\pi)^2}\CF
+\dots$ to obtain a correctly normalized current 
($\mu\frac{\partial}{\partial\mu}(\chi_A\za^{\ms})=0$).

\section{Ward identities}

A way to renormalize currents, especially also non-perturbatively,
is to enforce the continuum Ward identities 
\shortcite{Ward:1950xp}, \shortcite{Takahashi:1957xn} 
(in some cases only up to lattice artifacts), 
which for the case of flavor SU$(\Nf)\times$SU$(\Nf)$
are equivalent to current algebra in Minkowski space. General Ward identities
are obtained by making infinitesimal transformations in the functional
integral. For transformations which leave the functional measure invariant
we obtain relations of the form 
\be
\langle\delta S\,\calO\rangle=\langle\delta\calO\rangle\,.
\end{equation} 

For axial transformations (e.g. for $\Nf=2$) \shortcite{Luscher:1996sc} 
\be
\delta_A\psi(x)=\omega^a(x)\frac12\tau^a\gamma_5\psi(x)\,,\,\,\,\,
\delta_A\psibar(x)=\omega^a(x)\psibar(x)\frac12\tau^a\gamma_5\,,
\end{equation}
and working formally in the continuum (with an assumed chiral
invariant regularization) the action transforms as
\be
\delta_AS=\int_\mathcal{R}\rmd^4 x\,
\omega^a(x)\left[-\partial_\mu A_\mu^a(x)+2mP^a(x)\right]\,,
\end{equation}
where $P^a=\psibar\frac{\tau^a}{2}\gamma_5\psi$ is the pseudoscalar density
and we have assumed that $\omega^a(x)=0$ for $x$ outside a bounded region 
$\mathcal{R}$. For example if the observable $\calO=\calO_{\rm ext}$ 
has no support in $\mathcal{R}$ then the WI becomes 
\be
\langle\left[-\partial_\mu A^a_\mu(x)+2mP^a(x)\right]\calO_{\rm ext}\rangle=0\,,
\end{equation}
the famous {\it PCAC relation} which has many applications (see sect.~5.3).

If $\calO=\calO_{\rm int}\calO_{\rm ext}$, having support inside 
and outside $\mathcal{R}$, we obtain
\be
\int_\mathcal{R}\rmd^4x\,\omega^a(x)\langle\left[-\partial_\mu A^a_\mu(x)+2mP^a(x)\right]
\calO_{\rm int}\calO_{\rm ext}\rangle=
\langle\delta_A\calO_{\rm int}\calO_{\rm ext}\rangle\,,
\end{equation}
which in the limit $\omega^a(x)\to{\rm constant}$ inside $\mathcal{R}$, 
and $m=0$ simplifies to 
\be
\omega^a\int_{\partial\mathcal{R}}\rmd\sigma_\mu(x)\,
\langle A^a_\mu(x)\calO_{\rm int}\calO_{\rm ext}\rangle=
-\langle\delta_A\calO_{\rm int}\calO_{\rm ext}\rangle\,,
\end{equation}
where $\rmd\sigma_\mu$ is the outward normal to the boundary of $\mathcal{R}$.
For the case $\calO_{\rm int}=A^b_\nu(y)$ and $\mathcal{R}$ the
region between two fixed time hyperplanes at $t_2,t_1$,
(and using 
$\delta_A A_\nu^b(x)=-i\omega^a(x)\epsilon^{abc}V_\nu^c(x)$)
the WI reads ($t_2>t_1$),
\be
\int\rmd^3x\,\langle\left[A_0^a(\bfx,t_2)-A_0^a(\bfx,t_1)\right]
A_\nu^b(y)\calO_{\rm ext}\rangle
=i\epsilon^{abc}\langle V^c_\nu(y)\calO_{\rm ext}\rangle\,,
\end{equation}
 which is equivalent to the current algebra relation 
$\left[A^a_0(\bfx,t),A_0^b(\bfy,t)\right]
=i\delta^{(3)}(\bfx-\bfy)\epsilon^{abc}V_0^c(\bfx,t)$
in Minkowski space. All CA relations can be obtained analogously.

\section{Scale dependent renormalization}

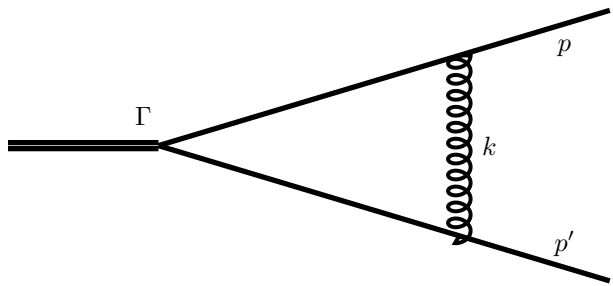
\begin{figure}
\hspace{2.5cm}
\psset{unit=2mm}
\begin{pspicture}(0,-10)(70,10)
\psset{linewidth=2pt}
\qline(0,0.2)(10,0.2)           
\qline(0,-0.2)(10,-0.2)           
\qline(10,0)(40,9)         
\qline(10,0)(40,-9)        
\rput(37,6.5){$p$}
\rput(37,-6.5){$p'$}
\rput(9,2){$\Gamma$}
\rput(32,0){$k$}
\pscoil[linewidth=1.5pt,coilarm=0,coilwidth=1.5]{-}(30,6)(30,-6)
\end{pspicture}
\caption{1-loop diagram contributing to a $\psibar\Gamma\psi$ vertex function} 
\label{figdensityvertex}
\end{figure}

Most composite operators require scale dependent renormalization. 
The non-perturbative
renormalization of such operators will be postponed to later sections.
Here we just outline the simple perturbative 1-loop computation 
of the anomalous dimensions of bilinear quark operators $\psibar\Gamma\psi$
(e.g. the pseudoscalar density $\Gamma=\gamma_5$ or $\Gamma=\tau^a/2$),
in the continuum MS scheme. At 1-loop the bare vertex function of 
the density with a quark-antiquark pair is given by the diagram with 
a gluon exchange in Fig.~\ref{figdensityvertex}: 
using dimensional regularization
\be
\Gamma(p,p')=-T^aT^a g^2\int\frac{\rmd^Dk}{(2\pi)^D}\,
\frac{\gamma_\nu\gamma(k+p)\Gamma\gamma(k-p')\gamma_\mu}{(k+p)^2(k+p')^2}
\frac{1}{k^2}
\left[\delta_{\mu\nu}-(1-\lambda^{-1})\frac{k_\mu k_\nu}{k^2}\right]
\,.
\label{Gammap}
\end{equation}
Noting that the integral is only logarithmically divergent for $D=4$,
for the computation of the UV divergent parts we can set $p=p'=0$
in the numerator. Without any previous experience of dimensional
regularization one can accept that the integral involved is singular
as $\epsilon\to0$,
\be
\int\frac{\rmd^Dk}{(2\pi)^D}F(k,p,p')\sim \frac{C}{\epsilon}\,,
\,\,\,\,
F=\frac{1}{(k+p)^2(k+p')^2}\,,\,\,\,\,C=\frac{1}{(4\pi)^2}\,,
\end{equation}
and it immediately follows for the integrals appearing in (\ref{Gammap}),
\ba
&&\int\frac{\rmd^Dk}{(2\pi)^D}\frac{k_\rho k_\tau}{k^2}F(k,p,p')
\sim \frac{C}{D\epsilon}\delta_{\rho\tau}\,,\,\,\,\,\
\\
&&\int\frac{\rmd^Dk}{(2\pi)^D}
\frac{k_\mu k_\nu k_\rho k_\tau}{(k^2)^2}F(k,p,p')
\sim \frac{C}{D(D+2)\epsilon}\left(\delta_{\mu\nu}\delta_{\rho\tau}
+2\,\,{\rm perms}\right)\,.\,\,\,\,\
\ea
Then we easily deduce
\be
\Gamma(p,p')_{\rm div}\sim - T^aT^a g^2\frac{C}{\epsilon}
\left[\frac{1}{D}\gamma_\mu\gamma_\rho\Gamma\gamma_\rho\gamma_\mu
-(1-\lambda^{-1})\Gamma\right]\,.
\end{equation}
For example
\footnote{problems with naive $\gamma_5$ algebra 
do not come in at this stage},
noting for our normalization $T^aT^a=-\CF$,
\ba
{\rm for}\,\,\,\Gamma=\gamma_\mu\gamma_5:
\,\,\,\,\Gamma(p,p')_{\rm div}&=& g^2\frac{C\CF}{\epsilon}
\gamma_\mu\gamma_5\left[\frac{(2-D)^2}{D}-(1-\lambda^{-1})\right]
\nonumber\\
&\simeq& g^2\frac{\CF}{(4\pi)^2\epsilon}\lambda^{-1}\gamma_\mu\gamma_5\,,
\\
{\rm for}\,\,\,\Gamma=\gamma_5:
\,\,\,\,\Gamma(p,p')_{\rm div}&=& g^2\frac{C\CF}{\epsilon}
\gamma_5\left[D-(1-\lambda^{-1})\right]
\nonumber\\
&\simeq& g^2\frac{\CF}{(4\pi)^2\epsilon}(3+\lambda^{-1})\gamma_5\,.
\ea
To complete the computation of the renormalization factor we have to compute
the quark field renormalization factor $Z_2$ to 1-loop. 
I leave this as an exercise;
but if we accept that the axial current doesn't need any divergent
renormalization to one loop this contribution must just cancel the
contribution above 
$Z_2^{(1)}=-\frac{\CF\lambda^{-1}}{(4\pi)^2}\frac{1}{\epsilon}$, 
and we deduce that the divergent part of the 
pseudoscalar density is given by
\be
\zp=1-\frac{3\CF}{(4\pi)^2}\frac{1}{\epsilon}g^2+\dots
\end{equation}
Note $\zp^{(1)}=-d_0/2$ as should be the case for $\zp\zm=1$.
For lattice regularization we would obtain $\zp=1+g_0^2 d_0\ln(a\mu)+\dots$
(in the MS scheme).

\section{Anomalies}

Any account on renormalization would not be complete without mentioning
anomalies. These involve symmetries present in the classical theory 
which are violated in the process of regularization
and which cannot be regained in the limit that the UV cutoff is removed.
This vast and important subject will be covered in the
lectures by Kaplan and thus no details will be given here. 
Let me just mention that we have already encountered one example which 
is that massless QCD breaks scale invariance at the quantum level. 
There is a mass parameter in the renormalized theory and the trace 
of the energy momentum tensor is non-zero
\be
\theta_{\mu\mu}=\frac{\beta(g)}{2g}N[F^2]\,.
\end{equation}  
For a discussion of the energy-momentum tensor in lattice gauge 
theories see \shortcite{Caracciolo:1991cp}.

Another famous example (see Kaplan's lectures) is the U(1) axial anomaly
\shortcite{Adler:1969gk} expressing the fact that the U(1) axial current 
is not conserved in the limit of massless quarks:  
\be
\partial_\mu A_{\rmR\mu}^0=2mN[\psibar\gamma_5\psi]+\frac{g^2}{32\pi^2}
N[\epsilon_{\mu\nu\rho\lambda}F_{\mu\nu}F_{\rho\lambda}]\,.
\end{equation}
This anomaly must be reproduced 
by a given formulation of lattice fermions 
in order that it can be considered an acceptable regularization of QCD.
For the many formulations available 
the way that this achieved varies quite considerably. For example
with Wilson fermions the measure is invariant under infinitesimal axial U(1)
transformations, but the Wilson term in the action breaks the symmetry 
and produces the correct anomaly in the continuum limit as was first shown 
by Karsten and Smit \citeyear{Karsten:1980wd}.
In the Ginsparg--Wilson formulation it is the measure which is not invariant
under the (modified) chiral U(1) symmetry transformations 
(whereas the action is); see sect.~6.

\section{Operator product expansions}

Consider correlation functions involving renormalized local gauge invariant
operators $\langle A(x)B(0)\phi(y_1)\dots\phi(y_r)\rangle$ with the $y_i$
physically separated from $x,0$ and from each other. In the limit $x\to0$
singularities appear which are described by local operators $\calO_n$ 
having the same global symmetries as the formally combined operator $AB$:
\be
A(x)B(0)\sim_{x\to0}\sum_n C_{AB}^{(n)}(x)\calO_n(0)\,,
\end{equation}
where the coefficients $C_{AB}^{(n)}(x)$ are c-numbers. This so called
{\it Wilson's operator product expansion} has been shown to hold in 
some generality in the framework of perturbation theory by Wilson and
Zimmermann \citeyear{Wilson:1972ee}. 
The relation is structural and thought to hold also at the 
non-perturbative level
\footnote{It has been shown non-perturbatively in some 2d models 
e.g. in the 2d Ising field theory \shortcite{Wu:1975mw} and
also in the massless Thirring model \shortcite{Wilson:1970pq}.}.
If the fields $A,B,\calO_n$ are multiplicatively renormalizable
then the coefficients obey the RG equation
\be
\left[\mu\frac{\partial}{\partial\mu}+\beta(g)\frac{\partial}{\partial g}
-\gamma_A-\gamma_B+\gamma_n\right]C_{AB}^{(n)}(x)=0\,.
\end{equation}
The engineering dimension of the coefficients are, in AF theories given by
those of the operators involved. A famous example is the OPE of
vector currents 
\be
V_\mu(x)V_\nu^\dagger(0)\sim_{x=0} C_{\mu\nu}^{(0)}(x)
+C_{\mu\nu}^{(1)}(x)N[F^2](0)+\dots
\end{equation}
The leading coefficient $C_{\mu\nu}^{(0)}(x)$ multiplying the identity
operator behaves like $1/(x^2)^3$ (for $x\to0$) up to logs, 
and (in the case of electromagnetic currents) 
describes the leading high energy behavior 
in $\rme^+\rme^-$ annihilation. It has been computed to high order in PT
\shortcite{Baikov:2009zz}. Because of gauge invariance
the next operators occurring here (in the massless theory) have dimension 4 
(only one term has been exhibited above) and so the corresponding 
coefficients e.g. $C_{\mu\nu}^{(1)}(x)\sim 1/x^2$. 
As stressed by many authors long ago e.g. by David \citeyear{David:1985xj},  
this does not mean that there are no terms in the 
associated physical amplitudes behaving like $1/(x^2)^2$. 
Unfortunately one still encounters the contrary 
statement in the literature; there is reference to ``the gluon condensate''
as a non-perturbative effect while forgetting that there can be 
non-perturbative effects in the coefficients $C_{\mu\nu}^{(n)}(x)$ and the
fact that to my knowledge there is at present no regularization 
independent definition of the gluon condensate! 
These considerations do of course not negate the strength of the OPE; 
PT gives the leading short distance behavior of the coefficients and the
sub-leading terms in the OPE give the leading effects in processes with 
non-vacuum external states. 

Another useful example of an application of the OPE is in non-leptonic decays 
in the framework of the Standard Model (e.g. $K\to2\pi$ 
\shortcite{Gaillard:1974nj}, \shortcite{Altarelli:1974exa} 
which will be 
discussed in detail in the lectures of Lellouch). I would just like to 
emphasize a few points below and in this discussion neglect quark masses. 
The typical Minkowski space amplitude 
for initial and final hadronic states $I,F$ has the form
\be
T_{FI}\propto\int\rmd^4x\,D(x,m_W)
\langle F|TJ_{1\mu}^L(x)J_{2\mu}^L(0)|I\rangle\,, 
\end{equation}
involving left-handed currents $J_1^L,J_2^L$, where $D(x,m_W)$ is the
scalar function occurring in the W-meson propagator.
Since the physical W-meson mass $m_W$ is much larger than typical strong 
interaction scales involved, short distances dominate the integral.
The simplest case to consider is that where the currents involve different
flavored quarks; in that case the OPE implies
\be
\int\rmd^4x\,D(x,m_W)TJ_{ru\mu}^L(x)J_{sv\mu}^L(0)\sim\sum_{\sigma=\pm}
h^\sigma(\mu/m_W,g)O^\sigma_{rsuv}(0)\,,
\end{equation}
with composite operators
\ba
O^\sigma_{rsuv}&=&O_{rsuv}\pm O_{rsvu}\,,
\\
O_{rsuv}&=&N\left[J_{ru\mu}^LJ_{sv\mu}^L\right]\,,\,\,\,\,
J_{ru\mu}^L=\psibar_r\gamma_\mu\frac12(1-\gamma_5)\psi_u\,.
\ea
The operators $O^\pm$ renormalize diagonally if one has a regularization 
preserving chiral symmetry
\footnote{Because of the difficulty with treating $\gamma_5$ 
in the framework of dimensional regularization, 
during the renormalization procedure one has to include mixing 
with ``evanescent operators" (ones which vanish for $D=4$).}. 
Restricting to that case we write the rhs in terms of the RGI
operators introduced in (\ref{RGIop}) as
\be
\sum_{\sigma=\pm}k^\sigma(m_W/\Lambda)O^{{\rm RGI}\sigma}_{rsuv}(0)\,,
\end{equation}
with
\ba
k^\sigma(m_W/\Lambda)&=&h^{\sigma}(1,\gbar^2(m_W))/C_\sigma(m_W/\Lambda)
\\
&=&(2b_0\gbar^2(m_W))^{-\gamma_\sigma^{(0)}/2b_0}
\left[1+k_{\sigma 1}\gbar^2(m_W)+\dots\right]\,.
\ea
The coefficients can be computed in PT e.g. $k_{\sigma1}=h_{\sigma1}
-(\gamma_\sigma^{(1)}-\gamma_\sigma^{(0)}b_1/b_0)/(2b_0)$, where $h_{\sigma1}$
is determined by 1-loop matching of the full amplitude with that of the OPE,
and $\gamma_\sigma^{(1)}$ is a two-loop anomalous dimension. 
There remains the important job of the lattice community to determine 
the non-perturbative amplitudes $\langle F|O^{{\rm RGI}\sigma}(0)|I\rangle$.

Wilson fermions break chiral symmetry and this has the effect,
as first pointed out by Martinelli \citeyear{Martinelli:1983ac}, 
that with this regularization the 
parity even part of $O^\pm$ mixes with other operators e.g. 
$O_\pm^{VA}\propto (VV-AA)\pm (u\leftrightarrow v)$. Although the number
of such operators is restricted by CPS symmetry (charge conjugation, parity,
and SU($\Nf$) in the limit $\mq=0$), there are still 3 such operators 
which makes Wilson fermions a rather awkward regularization for computing the
desired physical amplitudes in this case.

\chapter{Non-perturbative renormalization}

So far we have mainly considered renormalization in the framework of 
perturbation theory, which for QCD is only applicable to a class of high
energy processes. But QCD is a candidate theory for the hadronic interactions
at all energies. In particular in most numerical QCD computations we are
attempting to determine low energy observables. In such studies we fix 
the bare quark masses by fixing a sufficient number of scales 
e.g. $m_\pi/f_\pi\,,m_K/f_\pi\,..$ to their physical values, and then
$af_\pi(g_0)$ gives the lattice spacing $a(g_0)$ for the pion
decay constant $f_\pi$ fixed. This is called a 
{\it hadronic renormalization scheme}.

In order to connect a hadronic scheme to a perturbative scheme 
one could in principle proceed by computing a non-perturbatively
defined running coupling, e.g. using a 2-current vacuum correlation
function, over a large range of energies. Eventually at high energies 
(after taking the continuum limit) we can compare to perturbation theory
and estimate the scales e.g. $f_\pi/\Lambda_{\msbar}$. We can proceed
similarly for running masses and scale dependent renormalization constants.
However, despite the huge increase in available computational resources and
advances in algorithmic development,
to measure physical high energy $E$ observables with small lattice artifacts  
and negligible finite volume effects still raises the old practical problem
that the lattices required need too many points ($a\ll 1/E\ll 1/f_\pi<L$).

Many procedures have been applied in order to attempt to overcome this 
difficulty. The most naive way is to try to use the perturbative relation 
between the $\msbar$ coupling and the bare lattice coupling 
$\alpha_0=g_0^2/(4\pi)$ mentioned in sect.~2.2:
\ba
\alpha_\msbar(\mu)&=&\alpha_0+\alpha_0^2 d_1(a\mu)+\dots
\\
d_1(a\mu)&=&-8\pi b_0\ln(a\mu)+k\,,
\ea 
where the constant $k$ depends on the lattice action. 
As a non-perturbative input one computes a mass scale
e.g. a charmonium mass splitting, to give the lattice spacing $a(\alpha_0)$
in physical units. Now one can use the relation above to obtain an estimate  
for $\alpha_{\msbar}(\mu=s/a)$ (for some chosen factor $s$). This procedure
encounters many basic problems; firstly it is difficult to separate the
lattice (and finite volume) artifacts and estimate the systematic errors, 
and secondly for many actions (e.g. the standard plaquette action) 
one encounters large perturbative coefficients, 
in fact we have already seen this in the
large ratio between the lattice and $\msbar$ $\Lambda$-parameters. 
It was first observed by Parisi \citeyear{Parisi:1985iv}
\footnote{see also \shortcite{Martinelli:1982db}} 
that large contributions to this 
ratio come from tadpole diagrams, and that similar diagrams appear in the
computation of the average plaquette
\be
P=\frac{1}{N}\langle\tr\,U(p)\rangle=1+\rmO(g_0^2)\,.
\end{equation}
{\it Mean field improved bare PT} is an expansion in an alternative
bare coupling 
\be
\alpha_P\equiv\alpha_0/P\,. 
\end{equation}
If one now re-expresses $\alpha_{\msbar}$ in terms
of $\alpha_P$ one usually finds that the perturbative coefficients 
are reduced; e.g. choosing the scale $\mu$ such that the 1-loop term
is absent; for the standard action $N=3,\Nf=0$ one obtains
\be
\alpha_{\msbar}=\alpha_P+2.185\alpha_P^3+\dots\,,\,\,\,\,
{\rm for}\,\,\mu a=2.6\,,
\end{equation}
with a reasonably small 2-loop coefficient.
Computation of a mass scale and the plaquette expectation value
gives an estimate of the $\msbar$ coupling at $\mu=2.6/a$. There are
obviously many variants and the technique has been perfected by Lepage
and Mackenzie \citeyear{Lepage:1992xa}. 
But I think it is still true to say that 
systematic errors in these determinations are difficult to estimate 
and the scales
achieved are not so high that one can be confident to use the result
as initial conditions for running the RG equation with perturbative
beta-functions to higher energies.

\section{Intermediate regularization independent momentum scheme}

Another approach is to use an {\it intermediate regularization independent
momentum scheme} which will be mentioned again in sect.~5.3.
In this approach one usually
considers correlation functions involving some basic fields and thus one
has to fix a gauge (and tackle the Gribov ambiguity problem if a
covariant gauge is used). Using similar ideas non-perturbative 
running couplings can be defined as they are in PT 
e.g. from the 3-point gluon vertex function \shortcite{Alles:1996ka}.
For a covariant gauge the full gluon propagator in the 
continuum has the form (\ref{fullgprop}),
and the 3-gluon vertex function has the structure ($p_1+p_2+p_3=0$)
\be
\Gamma^{a_1a_2a_3}_{\mu_1\mu_2\mu_3}(p_1,p_2,p_3)=
-if^{a_1a_2a_3}F_{\mu_1\mu_2\mu_3}(p_1,p_2,p_3)
+id^{a_1a_2a_3}D_{\mu_1\mu_2\mu_3}(p_1,p_2,p_3)\,,
\end{equation}
where $f,d$ are the SU($N$) invariant tensors. At the ``symmetric point'' (SP)
$p_i^2=M^2\,,p_ip_j=-\frac12 M^2\,,i\ne j$ we have for the first amplitude
\be
F_{\mu_1\mu_2\mu_3}|_{\rm SP}=F(M^2)\left\{\delta_{\mu_1\mu_2}(p_1-p_2)_{\mu_3}
+2\,{\rm perms}\right\}+R_{\mu_1\mu_2\mu_3}
\end{equation}
where $n_{\mu_1}R_{\mu_1\mu_2\mu_3}=0$ if $np_i=0$ for all $p_i$. 
The symmetric point was also introduced by Lee and Zinn-Justin 
\citeyear{Lee:1972fj}
as a set of symmetric non-exceptional momenta in order to avoid potential 
IR problems (at least in the framework of PT). 
One can now define a renormalized running coupling by 
(\shortciteNP{Celmaster:1979km}, \citeyearNP{Celmaster:1979dm})
\be
g_{\rm MOM}(M^2)=F(M^2)\left[1+\Pi(M^2)\right]^{-3/2}\,,
\end{equation}
where $\Pi(k^2)$ is the dynamical function appearing in the
full gluon propagator (\ref{fullgprop}).

With an analogous construction for the lattice regularization,
one then measures $g_{\rm MOM}$ at various values of allowed 
$M^2=k\left(\frac{2\pi}{L}\right)^2$ 
\footnote{Note on a given $(L/a)^4$ lattice $k$ can only take certain 
allowed integer values.}
at a given value of the bare coupling $g_0$ at which one also 
measures a mass scale in order to specify $M^2$ in physical units. 
One then, if possible, repeats the procedure at other values of $g_0$
in order to attempt a continuum limit extrapolation. Then at the largest
values of $M^2$ one has to resort to perturbative evolution in order 
to reach high energies. The method avoids the use of bare PT, 
but for presently feasible lattice sizes one cannot really reach
sufficiently large $M^2$, and in the most optimistic case one
only has a small window of $M^2$ without too severe lattice artifacts.

\section{Recursive finite size technique} 

Lattice simulations are necessarily performed at finite volumes
and these effects are a source of systematic error if the measured 
physical quantities are desired in infinite volume. On the other hand
we can also make use of the finite volume as a probe of the system,
as developed in QFT to a high degree by L\"{u}scher 
(\citeyearNP{Luscher:1990ux}, \citeyearNP{Luscher:1991cf}).
For example in Aoki's lectures he described how one can extract 
infinite volume scattering data from measurements of finite volume 
effects on spectra at large volumes. 

Our interest here is to overcome
the renormalization problem mentioned above and for this purpose
it is useful to define a renormalized coupling depending on the
volume $\alphaFV$ e.g. one determined
in terms of the force between two static quarks
\be
\alpha_{q\bar{q}}(L)\propto\left\{r^2 F_{q\bar{q}}(r,L)\right\}_{r=L/2}\,.
\end{equation}
There are infinitely many acceptable choices  
and at large $L$ their behaviors can be completely different.
The important feature which characterizes them is that at small $L$
(where the spectral properties are vastly different from familiar
infinite volume spectra) we can use PT to compute them as a power
series in $\alpha_{\msbar}$ (starting linearly):
\be
\alphaFV(L)=\alpha_{\msbar}(\mu)+\alpha_{\msbar}(\mu)^2
\left[8\pi b_0\ln(\mu L)+c_{\FV}\right]+\dots
\end{equation}

Let us assume that we have decided on the definition of the 
finite size coupling.
The tactic to connect a hadronic scheme to a perturbative one, 
which goes under the name of {\it the recursive finite size technique}
\footnote{The RFST was first submitted by L\"{u}scher as an
(unaccepted) project proposal for the European Monte-Carlo 
Collaboration (${\rm EMC}^2$) formed after an early realization that 
large scale simulations required large collaborations.}, 
consists of the following steps. \\
1) Set the scale on the lattice with largest physical 
extent $L=\Lmax$ so that $\Lmax$ is known in physical units
say $\sim 0.5\fm$, and compute $\alphaFV(\Lmax)$. \\ 
2) Now perform non-perturbative evolution until $\alphaFV(L)$
is known on a lattice of much smaller size say $\sim 0.005\fm$. \\
3) Assuming perturbative evolution has apparently set in at the scale
reached in step 2, one continues with perturbative evolution and eventually
obtains the $\Lambda$--parameter in the FV scheme $\Lambda_{\mathrm{FV}}$
in physical units. \\
4) Relate the coupling to the $\msbar$-scheme to 1-loop
to obtain the ratio of $\Lambda$--parameters, and hence $\Lambda_{\msbar}$ 
in physical units. One can then use perturbative running to compute 
$\alpha_{\msbar}$ at any HE scale say $\mu=m_W$.

\begin{figure}
\psfig{figure=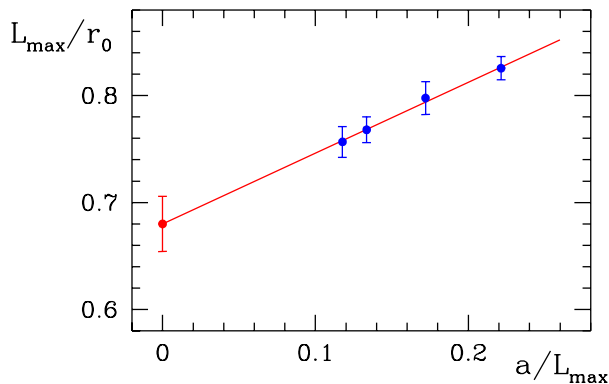,width=10cm}
\vspace{-2cm}
\caption{Extrapolation of $L_{\rm max}/r_0$ to the continuum limit}
\label{r0extrap}
\end{figure}

In the following we will outline some of the steps in more detail.
The order that some of the steps are carried out in practice is not fixed, 
but let us start with step 1. Let us define 
\be
\Lmax: {\rm volume\,\,where}\,\,\gbar^2_{\FV}(\Lmax)={\rm fixed\,\,value}
\,\,3.48\,, 
\end{equation}
the precise latter value is of course not essential but chosen after 
initial test runs to correspond to lattice sizes $\sim 1\fm$. We would 
like
to determine $\Lmax$ in terms of some physical unit. 
Considering first the pure Yang-Mills theory, a convenient quantity
is Sommer's scale \shortcite{Sommer:1993ce} $r_0$ defined by 
$r_0^2F_{q\bar{q}}(r_0)=1.65$,
which in phenomenological heavy quark (charmonium) potential models 
corresponds to a distance $\sim 0.5\fm$. To compute $\Lmax/r_0$
we first select a value of the bare coupling $g_0$ on a large lattice
(say $L/a\sim 48$) where one can measure $r_0/a$ accurately with negligible
finite volume effects. At the same $g_0$ one measures $\gbar^2_{\FV}$
on smaller lattices $L/a=6,8,\dots 16$ and obtains $\Lmax/a$ by interpolation,
and hence $(\Lmax/r_0)(g_0)$. 
Now the procedure is repeated at other values and subsequently
the data is extrapolated to the continuum limit using the 
theoretically expected form of the artifacts (discussed in the next section) 
as illustrated in Fig.~\ref{r0extrap}.

\begin{figure}
\psfig{figure=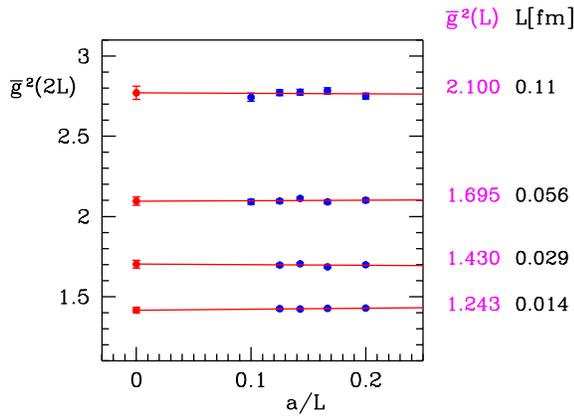,width=10cm}
\vspace{-2cm}
\caption{Extrapolation of $\Sigma$ to the continuum limit \`{a} la Symanzik}
\label{sigmaextrap}
\end{figure}

\begin{figure}
\psfig{figure=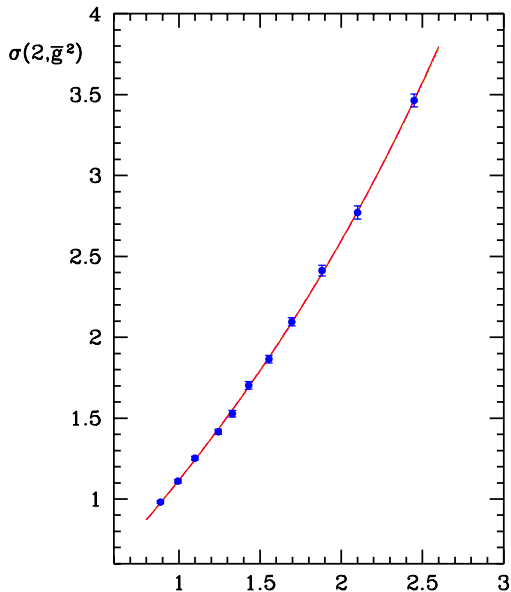,width=10cm}
\vspace{-2cm}
\caption{Results for the continuum step scaling function}
\label{sigmafit}
\end{figure}

Step 2, measuring the evolution was historically done in the reverse
order from the description above and the standard notation in the 
following is suited for this. In the continuum limit there is a 
well defined function $\sigma$, {\it the step scaling function}, 
relating the coupling at one volume to that at double the volume:
\be
\gbar_{\FV}^2(2L)=\sigma(\gbar_{\FV}^2(L))\,.
\end{equation}
With a lattice regularization this is modified to
\be
\gbar_{\FV}^2(2L)=\Sigma(\gbar_{\FV}^2(L),a/L)\,.
\end{equation}
One then starts with a convenient small lattice, say $L/a=8$,
and tunes $g_0$ such that $\gbar_{\FV}^2(L)$ equals some small value $u$.
At the same $g_0$ one computes $\gbar_{\FV}^2(2L)=\Sigma(u,a/L)$. 
One then repeats this for a manageable range of $a/L$ as illustrated
in Fig.~\ref{sigmaextrap}, and extrapolates the result to the continuum limit 
thus obtaining a value for $\sigma(u)=u'$. The whole procedure is then
repeated but this time starting with an initial value $\gbar_{\FV}^2(L)=u'$
(or a value close to $u'$). After this has been done many ($\rmO(10)$) times,
one ends up with a sequence of points for the continuum step scaling function
as illustrated in Fig.~\ref{sigmafit}:
\be
u_k=\sigma(u_{k+1})\,,\,\,\,\,u_0=3.48\,,
\end{equation}
giving $\gbar_{\FV}^2(L)$ at $L=2^{-k}\Lmax$. At the small values
of $L$ one can check whether $\sigma(u)$ is well described by the 
perturbative expectation
\be
\sigma(u)=u+(2\ln 2)b_0 u^2+\dots
\end{equation}
and if this is the case one can use the beta function
with perturbative coefficients to compute $\Lambda_{\FV}$ in physical units.
After that the steps are straightforward; a 1-loop computation relating
$\alpha_{\FV}$ to $\alpha_{\msbar}$ yields the ratio of $\Lambda$--parameters
and hence the desired value of the product $\Lambda_{\msbar}r_0$
relating the scales of the hadronic (LE) and perturbative (HE)
renormalization schemes.

It is clear that for the success of the RFS method described above, we
need a definition of the coupling which satisfies the following criteria:
a) it should be accurately measurable, b) it has preferably small lattice
artifacts, and c) it should be relatively easily computable in PT.

\section{The Schr\"{o}dinger functional}

After some extensive R\&D members of the Alpha Collaboration 
found that couplings based on the Schr\"{o}dinger functional (SF) 
\footnote{It was fortunate that M.~L\"{u}scher was already informed
about the SF in scalar theories \shortcite{Luscher:1985iu}, 
and that U.~Wolff had already
suggested a similar construction for the 2-d non-linear O($n$) sigma model.}
satisfy the above requirements \shortcite{Luscher:1992an}. 
It was further realized that the setup is 
also well suited for the computation of renormalization
constants in general, and that it is easily extended to include fermions
and to compute their running masses. 

\begin{figure}
\hspace{3cm}\psfig{figure=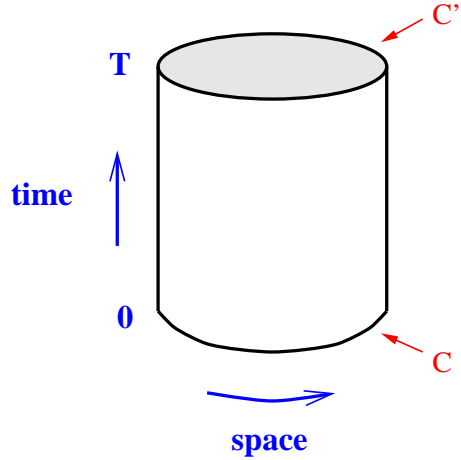,width=6cm}
\caption{Geometry of the Schr\"{o}dinger functional}
\label{schrodinger}
\end{figure}

In this framework one studies the
system in a cylindrical volume $\Lambda$ with Dirichlet boundary conditions in
one (the temporal) direction and periodic bc in the other (spatial)
directions (illustrated in Fig.~\ref{schrodinger}):
\be
A_k(\bfx,0)=C_k\,,\,\,\,\,
A_k(\bfx,T)=C'_k\,,\,\,\,\,
A_k(\bfx+L,t)=A_k(\bfx,t)\,.
\end{equation}
Formally in the continuum the SF is given by the functional integral
\be
Z(C,C')=\int_{SF\,\,{\rm bc}}D[A]\,\rme^{-S}\,,
\end{equation}
which is properly defined so that it is equal to the transition amplitude
$\langle C'|\rme^{-\hz T}\pz|C\rangle$ ($\hz$ the Hamiltonian, $\pz$ 
the projector on gauge invariant states) in the Hamiltonian formulation.

The renormalization of the SF in scalar field theories was first studied
by Symanzik \citeyear{Symanzik:1981wd}. He found that apart from the usual 
renormalization of the bare parameters and fields in the bulk 
one just requires some extra terms
on the boundaries, specifically spatial integrals over local fields of 
dimension $\le3$. L\"uscher's paper \citeyear{Luscher:1985iu} gives a 
clear introduction
to the subject. There are no such local gauge invariant operators in pure
Yang--Mills theory and so the (bare) SF should in this case not need any
renormalization besides the usual coupling constant renormalization.
We remark that one of the first papers considering the related topic of
the structure of Yang--Mills theories in the temporal gauge
were by Rossi and Testa 
(\citeyearNP{Rossi:1979jf}, \citeyearNP{Rossi:1980pg}).

For small bare coupling $g_0$ the functional integral is dominated by
fields around the absolute minimum of the action described by some
background field $B$. The SF then has a perturbative expansion
\be
-\ln Z(C,C')\sim \Gamma(B)=g_0^{-2}\Gamma_0(B)+\Gamma_1(B)+\dots
\end{equation}
If the boundary fields depend on a parameter $\eta$ then one can define
a renormalized running coupling as
\be
\gbar^2_{\rm SF}(L)=\left(\frac{\partial\Gamma_0(B)/\partial\eta}
{\partial\Gamma(B)/\partial\eta}\right)_{\eta=0,T=L}\,.
\end{equation}

Regularizing the gauge theory on the lattice the Scr\"{o}dinger functional
is an integral over all configurations of link matrices in ${\rm SU}(N)$:
\be
Z(C,C')=\int D[U]\rme^{-S[U]}\,,
\end{equation}
with the Haar measure and the $U$ satisfying periodic boundary conditions
in the spatial directions and Dirichlet bc in the time direction:
\be
U(x,k)|_{x_0=0}=W(x,k)\,,\,\,\,\,
U(x,k)|_{x_0=T}=W'(x,k)\,,\,\,\,\,k=1,2,3\,,
\end{equation}
where
\be
W(x,k)={\mathcal P}\exp\left\{a\int_0^1\rmd t\,
C_k\left(x+(1-t)a\hat{k}\right)\right\}\,,
\label{phasefactor}
\end{equation}
i.e. the SF is considered as a functional of the continuum fields $C,C'$
and the continuum limit $a\to0$ is taken with $C,C'$ fixed.

One can work in principle with any acceptable lattice action, 
the simplest being Wilson's plaquette action
\be
S=\frac{1}{g_0^2}\sum_p w(p)\Re\,\tr\,(1-U(p))\,,
\end{equation}
where the sum is over all plaquettes $p$ and the weight $w(p)=1$ except
for those lying on the boundary which is chosen $w(p)=1/2$ to 
avoid a classical $\rmO(a)$ effect.

Note that the derivative entering the definition of the coupling is 
\be
\frac{\partial\Gamma}{\partial\eta}=\left\langle
\frac{\partial S}{\partial\eta}\right\rangle\,.
\end{equation}
The expectation value appearing on the rhs involves only ``plaquettes" 
localized on the boundary. These are accurately measurable hence  
satisfying criteria (a) above.

As for the particular choice of the boundary fields $C,C'$ to make 
the perturbation expansion well defined we need the following
{\it stability condition}: if $V(x,\mu)=\exp\left[aB_\mu(x)\right]$
is a configuration of least action (with bc $C,C'$) then any other
gauge field with the same action is gauge equivalent to $V$.
Secondly we would like to have criterion (b). 
How to make optimal choices satisfying these demands is not at all obvious. 
Again after some experimentation the Alpha Collaboration made the choice 
of abelian bc e.g. for SU(3):
\be
C_k(x)=\frac{i}{L}
\left(\begin{array}{ccc}
\phi_1&0&0\\ 
0&\phi_2&0\\ 
0&0&\phi_3\\
\end{array}
\right)\,,
\,\,\,\,\sum_{\alpha=1}^3\phi_\alpha=0\,,\,\,\,
\phi_\alpha\,\,{\rm indep.\,\,of}\,\,x,k\,,
\end{equation}
and similarly for $C'$ involving elements $\phi'_i$.
The induced background field is abelian and given by ($T=L$)
\be
B_0=0\,,\,\,\,B_k(x)=C+(C'-C)x_0/L\,,\,\,\,\,C=C(\eta)\,.
\end{equation}
Stability has been proven \shortcite{Luscher:1992an} provided the 
$\phi$'s satisfy
\be 
\phi_1<\phi_2<\phi_3\,,\,\,\,\,|\phi_\alpha-\phi_\beta|<2\pi\,,
\nonumber
\end{equation}
(and similarly for $\phi'_\alpha$), 
and provided $TL/a^2$ is large enough, albeit the bound not being
very restrictive e.g. $TL/a^2>2\pi^2$ for $N=3$.

\begin{figure}
\psfig{figure=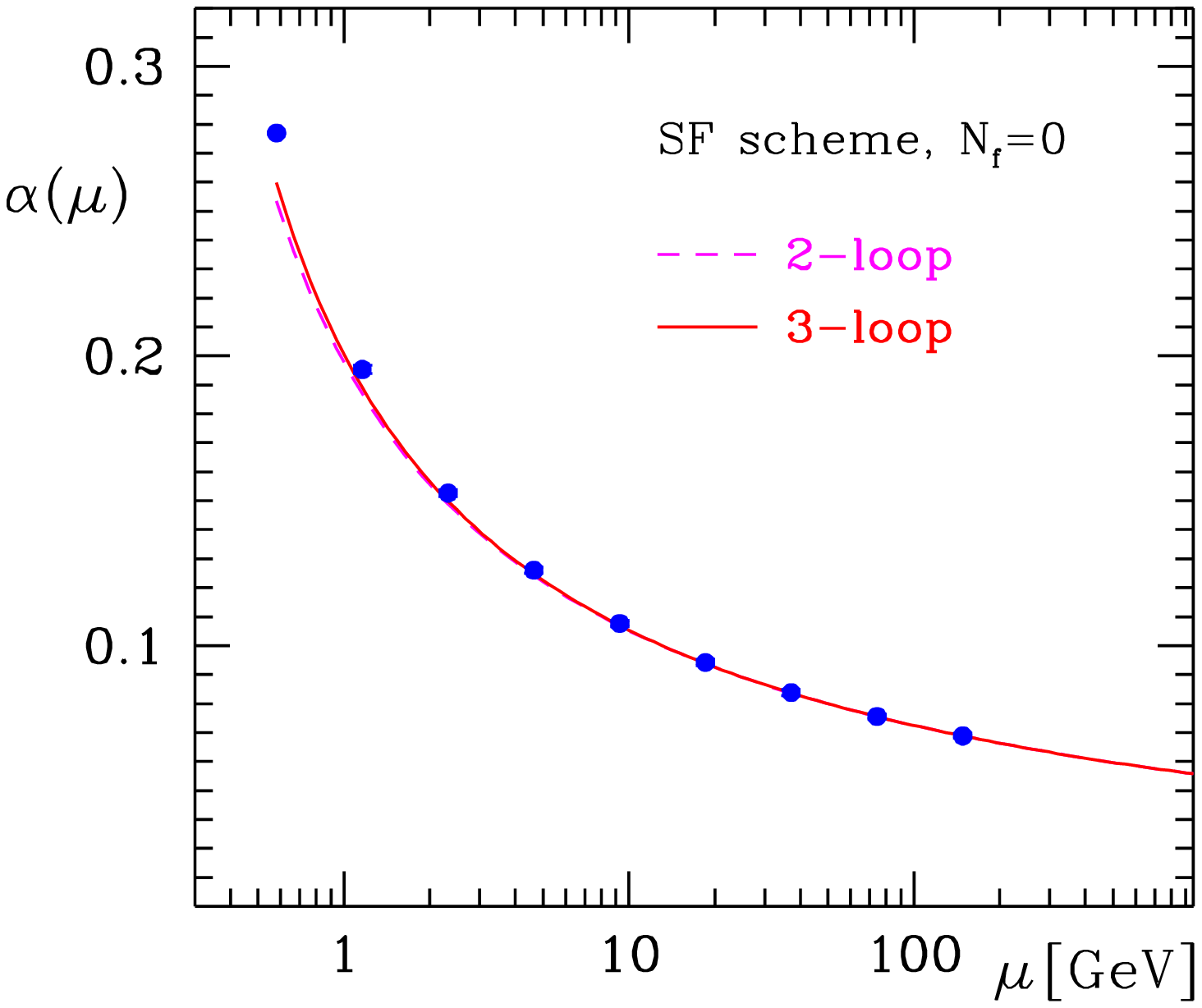,width=10cm}
\vspace{-2cm}
\caption{SF running coupling $\alpha(\mu)=\gbar^2_{\rm SF}(L)/4\pi\,,\,\,\mu=1/L$ ($\Nf=0$)}
\label{SFcoupling}
\end{figure}

With this setup the Alpha Collaboration produced measurements of a 
running coupling (in the continuum limit as far as it could be controlled) 
over a large range of energies
\footnote{This particular SF coupling runs similarly to PT down to
low energies, but this is not a universal property of non-perturbative 
running couplings.},
as depicted in Fig.~\ref{SFcoupling}.
At high energies the running is 
consistent with perturbative expectations, giving convincing numerical 
support to the (yet unproven) conventional wisdom that the critical 
coupling is $g_c=0$ and that the continuum limit of the lattice theory
is asymptotically free 
\footnote{Some additional evidence for the existence of  
non-perturbatively asymptotically free theories comes from studies of
integrable models in 2d (see e.g. \shortcite{Balog:2004mj}).}. 
Contrary to widespread opinion the latter property is non-trivial
(so far lacking rigorous proof) 
and some authors have questioned its validity \shortcite{Patrascioiu:2000mw}. 

\section{Inclusion of fermions}

The inclusion of fermions in the SF framework was first considered
by Sint (\citeyearNP{Sint:1993un}, \citeyearNP{Sint:1995rb}). In the 
continuum it has been argued by L\"{u}scher \citeyear{Luscher:2006df} 
that ``natural boundary conditions" involve linear conditions 
for the fields of lowest dimension. For Dirac fermions these take the form
\be
\mathcal{B}\psi|_{\rm bdy}=0
\end{equation}
where, in order to obtain a non-trivial propagator, 
the matrix $\mathcal{B}$ must not have maximal rank. 
If one demands invariance under space, parity,
time reflections ($x_0\to T-x_0$) and charge conjugation, one is left with 
the possibility 
\be
P_+\psi(x)=\psibar(x)P_-=0\,,\,\,\,x_0=0\,;\,\,\,
P_-\psi(x)=\psibar(x)P_+=0\,,\,\,\,x_0=T\,,
\label{SFfermionbc}
\end{equation}
(or $P_+\leftrightarrow P_-$) where $P_\pm=\frac12(1\pm\gamma_0)$.
Sint showed that with these homogeneous boundary conditions the SF
is renormalizable without the necessity of including any extra boundary terms.

With the lattice regularization where continuity considerations
are a priori missing, boundary conditions are implicit in the
specification of the dynamical fields (those to be integrated in the
functional integral) and the precise form of the action close 
to the boundary. For Wilson fermions the terms in the action
coupling close to the boundary e.g. near $x_0=0$ have the form
$\propto \psibar(a)P_+\psi(0)+\psibar(0)P_-\psi(a)$.
It is thus natural in the corresponding lattice SF to declare fields 
$\psi(x),\psibar(x)$ away from the boundary i.e. $0<x_0<T$ as the 
dynamical variables and expect that the bc's (\ref{SFfermionbc}) 
are recovered dynamically in the continuum limit. 
Often in the SF literature one sees the equations
\ba
P_+\psi(0,\bfx)&=&\rho(\bfx)\,,\,\,\,\,
\psibar(0,\bfx)P_-=\bar{\rho}(\bfx)\,,\,\,\,\,
\nonumber
\\
P_-\psi(T,\bfx)&=&\rho'(\bfx)\,,\,\,\,\,
\psibar(T,\bfx)P_+=\bar{\rho}'(\bfx)\,.
\nonumber
\ea
These are however not to be considered as specifying boundary conditions,
but describe couplings of sources for the undefined field components
near the boundary. For example, defining
\be
\xi(\bfx)=P_-\frac{\delta}{\delta\bar{\rho}(\bfx)}\,,\,\,\,
\bar{\xi}(\bfx)=-\frac{\delta}{\delta\rho(\bfx)}P_+\,,
\end{equation}
we can consider correlation functions of the form
\be
\langle\calO^a A^a(x)\rangle\sim\int[\rmd U\,\rmd\psi\,\rmd\psibar]\,
\calO^aA^a(x)\rme^{-S}\vert_{\rho=\bar{\rho}=\rho'=\bar{\rho}'=0}\,,
\end{equation}
where all sources are set to zero after differentiating and
\be
\calO^a\equiv-\sum_{\bfy,\bfz}\bar{\xi}(\bfy)\frac12\tau^a\gamma_5\xi(\bfz)\,.
\label{SFsource}
\end{equation}
In this setting the extra boundary counter-terms 
\footnote{of the form 
$\propto\int_{x_0=0}\rmd^3x\,\psibar(x)P_-\psi(x)
+\int_{x_0=T}\rmd^3x\,\psibar(x)P_+\psi(x)$}
appearing in Sint's original paper amount to a renormalization
of the sources $\xi(\bfx)\to Z^{1/2}_\xi\xi(\bfx)$.

An important point is that the SF fermion boundary conditions imply
a gap in the spectrum of the Dirac operator at least for $g_0$ small
enough. This has the consequence that simulations at zero quark mass 
$\mq=0$ with the Schr\"{o}dinger functional are not problematic.

Also an extra option is to impose quasi-periodic boundary conditions
in the spatial direction of the form
\be
\psi(x+L\hat{k})=\rme^{i\theta_k}\psi(x)\,,\,\,\,\,
\psibar(x+L\hat{k})=\rme^{-i\theta_k}\psibar(x)\,,
\end{equation}
which are equivalent to modifying the covariant derivative to
\be
(\nabla_k\psi)(x)=\frac{1}{a}\left[\rme^{ia\theta_k/L}U(x,k)\psi(x+a\hat{k})
-\psi(x)\right]\,,
\end{equation}
and similarly for $\nabla_k^*$. Such boundary conditions with various
choices of the $\theta_k$ serve as extra probes of the system.

\chapter{Lattice artifacts}

Probably in the future, computers will be so powerful that physically
large enough lattices will be measurable with very small lattice 
spacing $a$, such that lattice artifacts become numerically irrelevant.
Even so the question of the nature of lattice artifacts is of theoretical
interest. However, at present it is important in practice to gain insight 
in the form of the artifacts in order to make reliable extrapolations
of numerical data to the continuum limit.

\begin{figure}
\hspace{3cm}
\psfig{figure=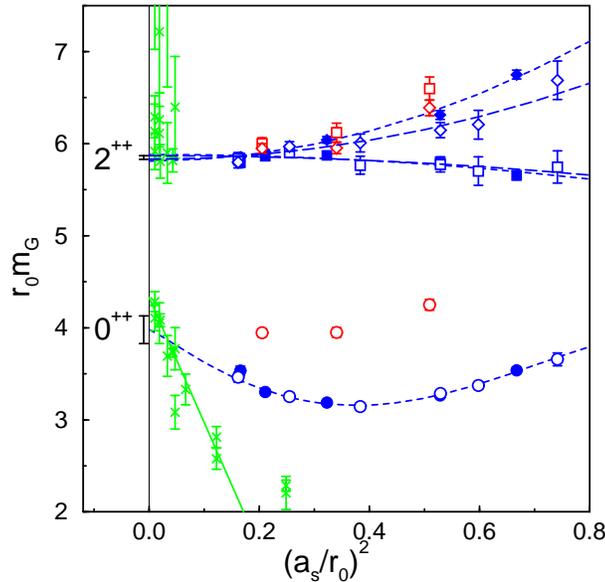,width=8cm}
\vspace{-0.5cm}
\caption{An old plot of glueball masses in pure YM measured with different 
actions \protect\shortcite{Morningstar:1996dn}; the results indicate 
universality. 
Green points (plaquette action) and the others from improved actions.}
\label{figglueball}
\end{figure}

Usually we make extrapolations of e.g. ratios of masses of the form
(\ref{artifacts}) assuming leading artifacts are predominantly polynomial
in the lattice spacing. Lattice artifacts are non-universal e.g. the
exponent $p$ and the coefficient $C_{12}$ in (\ref{artifacts}) depend
on the lattice action. This can be used in various ways e.g. if we 
simulate different actions and the data for glueball masses 
looks as in Fig.~\ref{figglueball},
this would be a support of the (expected) {\it universality} 
of the continuum limit and one could make a constrained joint fit.

More ambitious ways of using the non-universality involve designing actions
with larger values of the exponent $p$, which are called 
{\it Symanzik improved actions} 
(\shortciteNP{Symanzik:1979ph}, \citeyear{Symanzik:1981hc}) 
or even constructing {\it perfect actions} having in 
principle $p=\infty$ (see sect.~6).

Again most of our knowledge concerning lattice artifacts comes from studies
of perturbation theory. Some non-perturbative support for the validity of the
structure found there comes from investigations in the $1/n$-expansion of
QFT in 2 dimensions 
(see \shortcite{Knechtli:2005jh}, \shortcite{Wolff:2005nf}
and references therein), 
and also from many numerical simulations.

\section{Free fields}

Let us first consider a free scalar field theory on an infinite 
4--dimensional hyper-cubic lattice with standard action:
\be
S_0=a^4 \sum_{x,\mu}\frac12\left[\partial_\mu\phi(x)\partial_\mu\phi(x)
+m^2\phi(x)^2\right]\,,
\end{equation}
where $\partial_\mu f(x)=\left[f(x+a\hat{\mu})-f(x)\right]/a$. The 2--point 
function is
\be
\widetilde{G}(k)=\frac{1}{\hat{k}^2+m^2}\,,
\end{equation}
where $\hat{k}_\mu=\frac{2}{a}\sin\frac{ak_\mu}{2}$. 
Noting that for small $a$,
\be
\hat{k}^2=k^2-\frac{1}{12}a^2\sum_\mu k_\mu^4+\rmO(a^4)\,,
\end{equation}
we can write ($\phi_0$ the corresponding field in the continuum theory)
\be
\widetilde{G}(k)=\widetilde{G}_{\mathrm{cont}}(k)
-a^2\langle S_1^{\mathrm{eff}}\widetilde{\phi}_0(k)\phi_0(0)\rangle+\dots
\end{equation}
with
\be
S_1^{\mathrm{eff}}=-\int\rmd^4x\,\sum_\mu\frac{1}{24}
\partial_\mu^2\phi_0(x)\partial_\mu^2\phi_0(x)\,.
\end{equation}
On--shell information is obtained from the (lattice) two point function
$G(x)$ when $x$ is separated from 0 by a physical distance. Performing
the integral over $k_0$ we obtain the representation  
\be
G(x)=\int^{\pi/a}_{-\pi/a}\frac{\rmd^3k}{(2\pi)^3}\,\rme^{i\bfk\bfx}\,
\frac{\rme^{-\epsilon(\bfk,a,m)x_0}}{R(\bfk,m)}\,,
\end{equation}
where the energy spectrum $\epsilon(\bfk,a,m)$ is given by
\be
\cosh(a\epsilon(\bfk,a,m))-1=\frac12 a^2\left(\hat{\bfk}^2+m^2\right)\,.
\end{equation} 
Defining the pole mass by $m_{\mathrm{p}}=\epsilon(0,a,m)$ we obtain in the
continuum limit $a\to0$, $m_{\mathrm{p}}$ fixed: 
\be
\epsilon(\bfk,a,m)^2=m_{\mathrm{p}}^2+\bfk^2
-\frac{a^2}{12}T(\bfk,m_{\mathrm{p}})+\rmO(a^4)\,;
\,\,T(\bfk,m_{\mathrm{p}})
=\sum_{j}k_j^4+\bfk^2\left(\bfk^2+2m_{\mathrm{p}}^2\right)\,.
\end{equation}
The cutoff effects are (for $m_{\mathrm{p}}=0$) of order $>10\%$ for
$k>2\pi/(5a)$. One can improve this situation by adding $\rmO(a^2)$
terms to the action. This can be done in many ways, the simplest 
possibility being
\ba
S&=&S_0+cS_1\,,
\\
S_1&=&
a^4\sum_{x,\mu}\frac{a^2}{2}\partial_\mu^2\phi(x)\partial_\mu^2\phi(x)\,,
\ea
the latter involving interaction of next-to-nearest neighbors.
The energy spectrum is now given by the solution to
\be
\cosh(a\epsilon(\bfk,a,m))-1-2c\left[\cosh(a\epsilon(\bfk,a,m))-1\right]^2
=\frac12 a^2\left(\hat{\bfk}^2+m^2+ca^4\sum_j\hat{k}_j^4\right)\,.
\end{equation} 
Now for small lattice spacing $a$ the energy spectrum has the form
\be
\epsilon(\bfk,a,m)^2=m_{\mathrm{p}}^2+\bfk^2
+\left(c-\frac{1}{12}\right)a^2T(\bfk,m_{\mathrm{p}})+\rmO(a^4)\,,
\end{equation}
from which we see that the energy is $\rmO(a^2)$ improved 
if we chose $c=\frac{1}{12}$. 

Note that for the improved action another energy level is present but
its real part \footnote{ A spectral representation exists but energy levels
can be complex.}
always remains close to the cutoff, and hence it is irrelevant
for the continuum limit

\begin{exercise}
What is $S_1^{\mathrm{eff}}$ for the action 
$$
S_0+cS_1+d\sum_{x,\mu,\nu}\frac{a^2}{2}
\partial_\mu^*\partial_\mu\phi(x)\partial_\nu^*\partial_\nu\phi(x)\,.
$$
Show that the energy is improved for $c=\frac{1}{12}$ for arbitrary $d$.
\end{exercise}

\section{Symanzik's effective action}

Based on low order perturbative computations in various field theories 
one arrives at the following {\it conjecture}: In a large class of 
interacting lattice theories (in particular asymptotically free theories) 
there exists an (Symanzik's) effective continuum action
\be
S_1^{\mathrm{eff}}=\int\rmd^dy\,\mathcal{L}_1(y)\,,
\end{equation} 
such that a Green function of products of a multiplicatively 
renormalizable lattice field $\varphi$ at widely separated points $x_i$ 
takes the form
\ba
&&Z_\varphi^{r/2}\langle\varphi(x_1)\dots\varphi(x_r)\rangle_{\rm latt}
=\langle\varphi_0(x_1)\dots\varphi_0(x_r)\rangle_{\rm cont}
\nonumber\\
&&-a^p\int\rmd^dy\,
\langle\mathcal{L}_1\varphi_0(x_1)\dots\varphi_0(x_r)\rangle_{\rm cont}
\nonumber\\
&&+a^p\sum_{k=1}^r
\langle\varphi_0(x_1)\dots\varphi_1(x_k))\dots\varphi_0(x_r)\rangle_{\rm cont}
+\dots
\label{symaction}
\ea
where $\varphi_0,\varphi_1$ are renormalized continuum fields, in particular 
$\varphi_1$ is a sum of local operators of dimension $d_\varphi+p$
depending on the specific operator $\varphi$ and having the same lattice
quantum numbers as $\varphi$. The effective Lagrangian is a sum 
\be
\calL_1=\sum_i c_i(g(\mu),a\mu)\mathcal{O}_{\rmR i}(\mu)\,,
\label{L1eff}
\end{equation}
of local operators $\mathcal{O}_{\rmR i}$ of dimension $d+p$ having 
the symmetries of the lattice action.
$\mu$ is the renormalization scale e.g. of the dimensional regularization 
used in the continuum.

Note a) in the integral over $y$ one in general encounters singularities
at points $y=x_k$. A subtraction prescription must thus be applied, but
the arbitrariness in this procedure amounts to a redefinition of $\varphi_1$.
b) The coefficients $c_i$ are, as indicated, functions of the lattice 
spacing $a$, but the dependence is thought to be weak (logarithmic).

If the conjecture is true then one generically expects $\rmO(a^2)$
artifacts in pure Yang--Mills theory and $\rmO(a)$ effects with
Wilson fermions. All present numerical data seems consistent with
these expectations but until now only a small range of $a$ is available.

\section{Logarithmic corrections to $\rmO(a^2)$ lattice artifacts}

\begin{figure}
\begin{center}
\psfig{figure=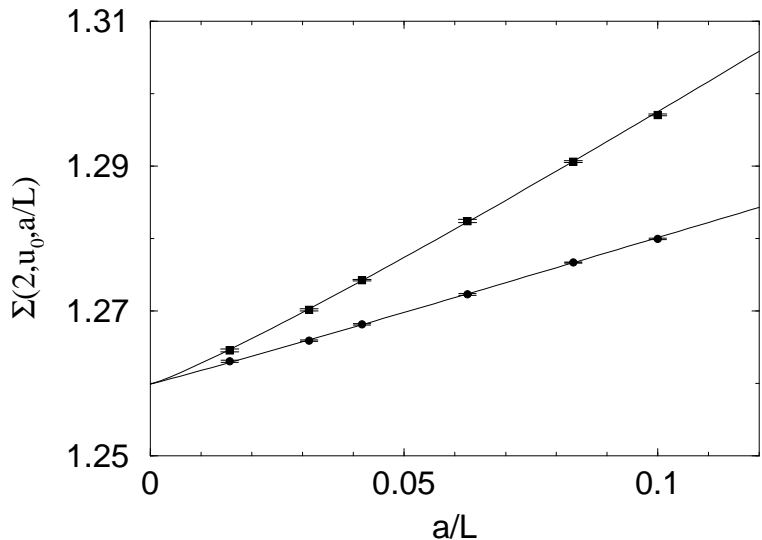,width=10cm}
\end{center}
\vspace{-0.5cm}
\caption{Monte Carlo measurements of the O(3) $\sigma$--model step 
scaling function at $u_0=1.0595$ for two lattice actions. 
The fits shown contain $a$ and $a\ln a$ terms. Fits of the 
form $k_1 a^2 +k_2 a^2\ln a + k_3 a^4$ have unacceptably 
high $\chi^2/{\rm dof}$.
}
\label{lww}
\end{figure}

In 2d lattice models e.g. the non-linear O($n$) sigma model,
which is perturbatively asymptotically free, one can simulate lattices
with very large correlation lengths ($>200a$). In these theories the
expectation is also $\rmO(a^2)$ artifacts. Hence it came as a surprise,
as mentioned by Hasenfratz in his LATT2001 plenary talk 
\shortcite{Hasenfratz:2001bz},
that data on a step scaling function in this model
seemed to show $\rmO(a)$ effects as illustrated in Fig.~\ref{lww}!
This was rather unsettling and motivated Balog, Niedermayer and myself
(\citeyearNP{Balog:2009np}, \citeyearNP{Balog:2009yj})
to investigate the logarithmic corrections to the $\rmO(a^2)$
in the framework of renormalized perturbation theory. We found that
generic artifacts in the $\rmO(n)$ sigma model are of the form 
$a^2\ln^s(a^2)$ with $s=n/(n-2)$. For $n=1$ the exponent is $s=3$, and 
such strong logarithmic corrections to the $\rmO(a^2)$
effects can explain the peculiar behavior, and yield good fits 
of the data for various actions \shortcite{Balog:2009np}. 
For $n=\infty$ the exponent is $s=1$ which is consistent with what 
is found in leading orders of the $1/n$ expansion
\shortcite{Wolff:2005nf}.

The steps involved in obtaining the result above are as follows.

1) Classify operators of dimension 4 (recall for this case $d=2$ and $p=2$)
which appear in the Symanzik effective Lagrangian (\ref{L1eff}).

2) Compute the $c_i$ at tree level (the coefficients normalized such that
$c_i=c_i^{(0)}+\rmO(g^2)$). Although finally interested only in 
on--shell observables, it is sometimes convenient to work off shell
and compute a sufficient number of correlation functions $G^{(r)}$
with a product of $r$ basic fields, in the continuum and on the lattice.
For the lattice Green function,
\be
G^{(r)}_{\rm latt}=G^{(r)}_{\rm cont}+a^2\sum_i c_i(g)G^{(r)}_i+\dots
\end{equation}
where $G_{i}^{(r)}$ are continuum correlation functions involving 
additional insertion of a composite field $\rmO_{\rmR i}$. 

3)  The ratios 
$\delta^{(r)}_i=G^{(r)}_i/G^{(r)}$ which characterize the lattice artifacts
(but are themselves independent of the lattice regularization) 
obey an RG equation of the form
\be
\left\{
\left(-a\frac{\partial}{\partial a}+\beta\frac{\partial}{\partial g}\right)
\delta_{ij}+\nu_{ij}\right\}\delta^{(r)}_j=0\,,
\end{equation}
where $\nu_{ij}$ is obtained from the mixing of the $\rmO_{\rmR i}$ to 1-loop
(see sect.~2.3).
If we have a basis where the renormalization is diagonal to one loop
$\nu_{ij}=-2b_0\triangle_i\delta_{ij}g^2+\dots$ then the operator 
associated to the largest value of $\triangle_i$ generically dominates 
the artifacts if the corresponding tree level coefficient $c_i(0)\ne0$.

The program should be carried out for lattice actions used for 
large scale simulations of QCD, when technically possible, in order 
to check if potentially large logarithmic corrections to lattice 
artifacts predicted by perturbative analysis appear.

\section{Symanzik improved lattice actions}

If Symanzik's conjecture is true it practically follows that it is 
possible to find a Symanzik improved lattice action such that 
$S_1^{\rm eff}=0$, 
i.e. for this action there are no lattice artifacts $\rmO(a^p)$. 
The conjecture is generally accepted for AF theories, but I should mention 
that a rigorous proof of the existence of a Symanzik improved lattice action 
for any  theory (including $\phi^4$) to all orders PT 
\footnote{Symanzik's published papers dealt with lower orders PT
probably he considered the generalization straight forward.}
is, to my knowledge, not complete. But there is an all order proof by
Keller \citeyear{Keller:1992xm} for the existence of Symanzik improved 
actions for $\phi^4_4$ and QED in the framework of a 
continuum regularization (using flow equations)
\footnote{Here the improvement refers to effects involving the 
UV cutoff $\Lambda$ occurring in the definition of the bare propagators.}. 
A subtle point is that the continuum limit of lattice $\phi^4_4$
theory is probably trivial i.e. a free theory. The renormalized coupling
goes to zero as $c/\ln(a\mu)$ and hence the continuum limit is
actually reached only logarithmically! 
Treating the renormalized coupling $g$ effectively as a constant for a range 
of cutoffs one has for small $g$ a perturbative Lagrangian description of 
the low energy physics, and in this case the Symanzik effective Lagrangian
describes the leading cutoff corrections to this.

An important ingredient of a lattice proof (to all orders PT) 
would presumably need a proof 
of the small $a$ expansion of an arbitrary $\ell$--loop Feynman diagram of
the form
\be
F(p,a)\sim a^{-\omega}\sum_{n=0}^\infty a^n
\sum_{r=0}^\ell (\ln am)^rF_{nr}(p)\,,
\label{lexpansion}
\end{equation}
which we are quite confident holds and hence often stated in the literature,
but which again has not, to my knowledge, been proven for $\ell\ge2$.

\begin{exercise}
As an example of an expansion of the form
(\ref{lexpansion}), consider a lattice $\phi^4_4$ theory 
with free propagator $1/[R(k,a)+m^2]$ with
$$
R(q/a,a)=a^{-2}r(q)\,,\,\,\,\,r(q)\sim_{q=0} q^2+c(q^2)^2+d\sum_\mu q_\mu^4+\dots
$$
Show that the tadpole integral
$J_1=\int_{-\pi/a}^{\pi/a}\rmd^4k\,[R(k,a)+m^2]^{-1}$ has an expansion of the
form
$$
J_1\sim a^{-2}\left\{r_0+a^2m^2[r_1+s_1\ln(am)]
+a^4m^4[r_2+s_2\ln(am)]
+\rmO(a^6)\right\}
$$
with $s_2=0$ if $c=d=0$.
\end{exercise}

\section{On--shell improved action for pure Yang--Mills theory}

\begin{figure}
\hspace{4.0cm}
\psset{unit=2mm}
\begin{pspicture}(0,-18)(26,6)
\psset{linewidth=2pt}
\psline{->}(0,0)(6,0)
\psline{->}(6,0)(6,6)
\qline(0,0)(0,6)
\qline(0,6)(6,6)
\psline{->}(14,0)(20,0)
\psline{->}(20,0)(26,0)
\psline{->}(26,0)(26,6)
\qline(14,0)(14,6)
\qline(14,6)(26,6)
\psset{linewidth=0.6pt}
\qline(20,0)(20,6)
\rput(3.5,-2.5){$(0)$}
\rput(20,-2.5){$(1)$}
\psset{linewidth=2pt}
\psline{->}(0,-7)(0,-13)
\psline{->}(0,-13)(4.9,-15)
\psline{->}(4.9,-15)(9,-13)
\psline{->}(9,-13)(9,-7)
\psline{->}(9,-7)(4.1,-5)
\psline{->}(4.1,-5)(0,-7)
\psset{linewidth=0.6pt}
\qline(4.9,-15)(4.9,-9)
\qline(4.9,-9)(0,-7)
\qline(4.9,-9)(9,-7)
\qline(0,-13)(4.1,-11)
\qline(4.1,-11)(9,-13)
\qline(4.1,-11)(4.1,-5)
\psset{linewidth=2pt}
\psline{->}(20,-13)(26,-13)
\psline{->}(26,-13)(26,-7)
\qline(20,-7)(26,-7)
\psline{->}(16,-15)(20,-13)
\qline(16,-15)(16,-9)
\qline(16,-9)(20,-7)
\psset{linewidth=0.6pt}
\qline(20,-13)(20,-7)
\rput(3.5,-17.5){$(2)$}
\rput(20.0,-17.5){$(3)$}
\end{pspicture}
\caption{4 and 6-link closed curves on the lattice}
\label{sixlinks}
\end{figure}
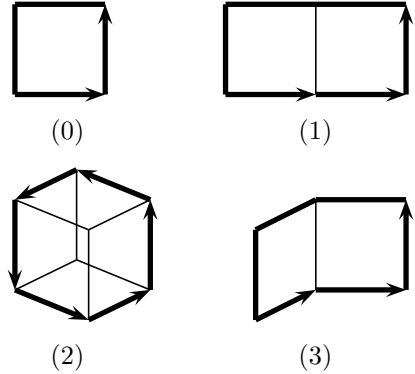

With the insight gained from our previous discussion we are now prepared 
to consider
improved actions for Yang--Mills theory in 4 dimensions
\shortcite{Luscher:1984zf}. As by now familiar,
the first step is to classify the independent (up to total derivatives)
gauge invariant operators of dimension 6, which are scalars under lattice 
rotations; there are three such operators \shortcite{Weisz:1982zw}:
\be
O_1=\sum_{\mu,\nu}\tr D_\mu F_{\mu\nu}D_\mu F_{\mu\nu}\,,\,\,
O_2=\sum_{\mu,\nu,\rho}\tr D_\mu F_{\nu\rho}D_\mu F_{\nu\rho}\,,\,\,
O_3=\sum_{\mu,\nu,\rho}\tr D_\mu F_{\mu\rho}D_\nu F_{\nu\rho}\,,
\end{equation}
where $D_\mu F_{\nu\rho}=\partial_\mu F_{\nu\rho}+g_0\left[A_\mu,F_{\nu\rho}\right]$.
Candidates for Symanzik improved actions have the form
\ba
S_{\rm imp}&=&\frac{2}{g_0^2}\sum_i c_i(g_0)
\sum_{\calC_i\in\calS_i}\calL(\calC_i)\,,  
\label{Simp}
\\
\calL(\calC)&=&\Re\,\tr\,[1-U(\calC)]\,,
\ea
where $U(\calC)$ is the ordered product of link variables around 
the closed curves $\calC$, and $\calS_i$ are sets with a given topology
e.g. 
$\calS_0=$ the set of plaquettes,\\
$\calS_1=$ the set of $2\times 1$ rectangles,
$\calS_2=$ the set of ``twisted chairs'',\\
$\calS_3=$ the set of ``chairs'',
as depicted in Fig.~\ref{sixlinks}.\\ 
Identifying the $U(x,\mu)$ with phase factors in the continuum
associated with the links as in (\ref{phasefactor}) 
(with $C_k$ replaced by $g_0A_\mu$), 
the classical small $a$ expansion of the local lattice operators is given by
\footnote{For this computation it is convenient to chose an axial gauge,
and it is sufficient to consider Abelian fields.}
\ba
\calO_i(x)&\equiv&\frac14\sum_{\calC\in\calS_i,x\in\calC}\calL(\calC)
\\
&=&a^4z_i \tr F_{\mu\nu}F_{\mu\nu}+a^6\sum_{j=1}^3 p_{ij}O_j(x)\,. 
\ea
We need only 4 lattice operators to represent the 4 continuum operators
of dimension 4,6 appearing in the effective action. We chose 
the sets of curves with smallest perimeter $\le6$ mentioned above,
but many other choices are admissible (and have appeared in the literature).
In order that the coefficient of $F^2$ in the classical expansion 
has the usual normalization the coefficients must satisfy
\be
c_0(g_0)+8c_1(g_0)+8c_2(g_0)+16c_3(g_0)=1\,.
\end{equation}
Further one finds that improvement of the classical action requires
\be
c_1(0)=-\frac{1}{12}\,,\,\,\,\,\,c_2(0)=c_3(0)=0\,.
\end{equation}
On the other hand improvement of on-shell quantities only requires 
two conditions among the coefficients. e.g. for the static 2-quark potential
at tree level one finds
\be
V(R)=-C\frac{g_0^2}{4\pi R}\left[1
+3\left(c_1(0)-c_2(0)-c_3(0)-\frac{1}{12}\right)\frac{a^2}{R^2}+\dots\right]\,. 
\end{equation}
Spectral quantities in finite volumes, e.g. with twisted boundary conditions,
can also be computed, and reproduces the improvement condition above 
and in addition the condition $c_2(0)=0$ 
(\shortciteNP{Luscher:1984xn}, \citeyearNP{Luscher:1984xo}). 
There are no other independent relations. This is as expected
because on-shell the operator $O_3$ can be dropped from the effective 
action since it vanishes when using the equations of motion. 
For on-shell improvement we conclude that we can choose
\be
c_3(g_0)=0\,\,\,\,\,{\rm for\,\,all\,\,} g_0\,.
\end{equation}
The coefficients $c_i$ can be computed to higher orders in PT. To 1-loop
they have been computed from the same observables as mentioned above 
\shortcite{Luscher:1985zq}.

In principle the $c_i(g_0)$ could be computed non-perturbatively  
by demanding cutoff effects to vanish from some spectral levels e.g.
requiring the $J^{\rm PC}=2^{++}$ states to be degenerate. This has
not been done yet, the reason being that the physical goal is QCD and
to achieve full $\rmO(a^2)$ improvement in that theory is much more 
difficult because
there are more dimension $6$ operators to be taken into account, in
particular those involving products of 4 quark fields. Never the less
improved gauge actions, based on various considerations, are used 
in practical large scale numerical simulations.

\chapter{$\rmO(a)$ improved Wilson fermions}

\begin{figure}
\hspace{3cm}$\widehat{F}_{\mu\nu}(x)\,\,\,\,\propto\,\,\,$
\psfig{figure=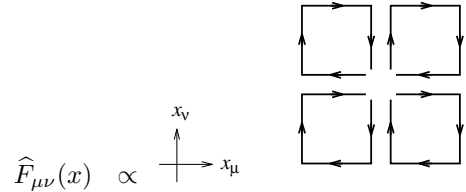,width=4cm}
\caption{Lattice representative of $g_0F_{\mu\nu}$}
\label{clover}
\end{figure}

As discussed in Hernandez' lectures, there are many ways of putting
fermions on the lattice, each having their own particular advantages
and disadvantages. In this section I will only discuss lattice artifacts
with Wilson fermions. From the tree level coupling of quarks to one gluon 
\be
\sim g_0 T^a\left\{\gamma_\mu-\frac12 a(p+p')_\mu+\rmO(a^2)\right\}\,,
\end{equation}
we immediately see that there are $\rmO(a)$ off--shell effects, and some
persist on--shell. The on-shell $\rmO(a)$ improvement program 
\shortcite{Luscher:1996ug} 
proceeds on the same lines as for the pure gauge fields in the last section.
The first step is thus to classify the independent local gauge invariant 
operators of dimension 5 that can occur in the effective Lagrangian $\calL_1$:
\ba
O_1&=&g_0\psibar i\sigma_{\mu\nu} F_{\mu\nu}\psi\,,\,\,\,\,
\sigma_{\mu\nu}=\frac{i}{2}\left[\gamma_\mu,\gamma_\nu\right]\,,\,\,\,
\\
O_2&=&\psibar D_\mu D_\mu\psi
+\psibar\overleftarrow{D}_\mu\overleftarrow{D}_\mu\psi
\,,\,\,\,\,\,\,\psibar\overleftarrow{D}_\mu
=\psibar\left(\overleftarrow{\partial}_\mu-g_0A_\mu\right)\,,
\\
O_3&=&mg_0^2\tr F_{\mu\nu}F_{\mu\nu}\,,\,\,\,\,
O_4=m\psibar\left(D_\mu-\overleftarrow{D}_\mu\right)\gamma_\mu\psi\,,\,\,\,
O_5=m^2\psibar\psi\,.
\ea
On-shell we can use the equations of motion $(\gamma D+m)\psi=0$
to derive relations
\be
O_1-O_2+2O_5\simeq0\,,\,\,\,\,
O_4+2O_5\simeq0\,,\,\,\,\,
\end{equation}
which can be used to eliminate $O_2,O_4$. A Symanzik--improved action
should then be constructible by adding a linear combination 
of lattice representations of $O_1,O_3,O_5$ 
\be
\delta S=a^5\sum_x\left\{c_1(g_0)\widehat{O}_1(x)
+c_3(g_0)\widehat{O}_3(x)+c_5(g_0)\widehat{O}_5(x)\right\}\,,
\end{equation}
to the Wilson fermion action. Now $\widehat{O}_3,\widehat{O}_5$
are already present in the original lattice action, so adding these terms
merely corresponds to a rescaling of the bare coupling and masses by terms 
$\sim 1+\rmO(am)$. i.e. they can be dropped until we discuss renormalization.
We conclude that the on-shell improved Wilson action has only one extra term
\be
S_{\rm imp}=S_W+a^5\sum_x c_{\rm SW}(g_0)\psibar(x)\frac{i}{4}
\sigma_{\mu\nu} \widehat{F}_{\mu\nu}(x)\psi(x)\,,
\end{equation}
where the lattice representative $\widehat{F}_{\mu\nu}(x)$ of 
$g_0F_{\mu\nu}(x)$ depicted in Fig.~\ref{clover} has a ``clover-leaf form" . 
This action was first written down by Sheikholeslami and Wohlert 
\citeyear{Sheikholeslami:1985ij}. 
The coefficient $c_{\rm SW}(g_0)$
is known to 1-,2-loop order of PT for various gauge actions.

Two comments are in order here. Firstly as for any improvement,
$\rmO(a)$ improvement is only an asymptotic concept; it could happen
that at some $g_0$ we have bad luck and $\rmO(a^2)$ effects are bigger than
$\rmO(a)$ effects for some quantities. 
Secondly, a rather nice feature is that the 
extra overhead CPU cost in simulations of adding a SW term is not very
substantial.

\section{$\rmO(a)$ improvement of operators}

In order to eliminate lattice artifacts in correlation functions involving
composite operators, we must also improve the operators themselves, by
adding local terms of higher dimension having the same quantum numbers.
For example for the axial iso-vector current $A_\mu^a$ the dimension 4 operator
$(A_\mu^a)_1$ in the effective Lagrangian description could involve
terms
\ba
O^a_{1\mu}&=&\psibar\left(D_\nu-\overleftarrow{D}_\nu\right)
\sigma_{\mu\nu}\gamma_5\frac12\tau^a\psi\,,
\\
O^a_{2\mu}&=&\partial_\mu\psibar\gamma_5\frac12\tau^a\psi=\partial_\mu P^a\,,
\,\,\,\,\,\,\,O^a_{3\mu}=mA_\mu^a\,.
\ea
On-shell one can eliminate $O^a_{1\mu}$ in favor of $O^a_{2\mu},O^a_{3\mu}$;
moreover $O^a_{3\mu}$ is just a renormalization of the original 
operator. An ansatz for an improved lattice bare operator is then
\be
\left(A_I^a\right)_\mu=A^a_\mu+a\ca\frac12
\left(\partial_\mu+\partial_\mu^*\right)P^a\,,
\end{equation}
and similarly for other operators e.g.
\be
P_I^a=P^a\,,\,\,\,\,
\left(V_I^a\right)_\mu=V^a_\mu+a\cv\frac12
\left(\partial_\nu+\partial_\nu^*\right)i\psibar\sigma_{\mu\nu}
\frac12\tau^a\psi\,.
\end{equation}
The improvement coefficients $\cv,\ca$ appearing above are $\rmO(g_0^2)$.

\section{Mass independent renormalization scheme}

In lattice QCD it is often advantageous to use a mass independent 
renormalization scheme as discussed in sect.~2. Without $\rmO(a)$ 
improvement this would involve 
renormalization constants independent of the quark masses and take the form
\ba
\gr^2&=&g_0^2 Z_g(g_0^2,a\mu)\,,
\\
\mr&=&\mq Z_m(g_0^2,a\mu)\,,\,\,\,\,\mq=m_0-\mc\,.
\ea
To obtain correlation functions at physical distances to approach their
continuum limit at a rate $\rmO(a^2)$ with improved Wilson fermions, we 
must modify the form of the renormalized parameters (and thereby account
for the terms $\widehat{O}_3,\widehat{O}_5$ we dropped previously)
in order to avoid uncanceled $\rmO(a\mq)$ effects 
\shortcite{Luscher:1996ug}.
We do this by introducing modified bare parameters
\ba
\widetilde{g}_0^2&=&g_0^2\left(1+\bg a\mq\right)\,,
\\
\widetilde{m}_{\rm q}^2&=&\mq\left(1+\bm a\mq\right)\,,
\ea
and define renormalized parameters through
\ba
\gr^2&=&\widetilde{g}_0^2 Z_g(\widetilde{g}_0^2,a\mu)\,,
\\
\mr&=&\widetilde{m}_{\rm q} Z_m(\widetilde{g}_0^2,a\mu)\,.
\ea
We can compute the new improvement coefficients in perturbation theory.
e.g. from the ``pole mass" at tree level
\be
m_{\rm P}=\frac{1}{a}\ln(1+am_0)=\mq-\frac12 a\mq+\dots\,,
\end{equation}
we deduce 
\be
\bm=-\frac12+\rmO(g_0^2)\,.
\end{equation} 
The lowest contribution to the coefficient $\bg$ is $\rmO(g_0^2)$ and
first obtained by Sint and Sommer \citeyear{Sint:1995ch} from the 1-loop 
computation of the SF coupling with fermions at fixed $z=\mq L$:
\be
\gbar_{\rm SF}^2(L)=g_0^2+g_0^4\left[2b_0\ln(L/a)+C(z)+\Nf ka\mq
+\dots\right]\,,
\end{equation} 
from which one deduces $\bg=\Nf k g_0^2+\dots$.

Similar factors are required for renormalized composite operators
\ba
A_{{\rm R}\mu}^a&=&\za\left(1+\bA a\mq\right)(A_I^a)_\mu\,,
\\
P_{\rm R}^a&=&\zp\left(1+\bp a\mq\right)P^a\,.
\ea
The coefficients $\bA,\bp=1+\rmO(g_0^2)$, which were first computed 
to 1-loop order by \shortcite{Gabrielli:1990us}, 
don't depend on renormalization 
conditions - the latter are applied at $\mq=0$.

\section{Determination of RGI masses and running masses}

We recall from sect.~2 that renormalization group invariant masses
are scheme independent. Running quark masses can be defined 
non-perturbatively in various ways; a most efficient way makes use
of the PCAC relation:
\be
\mbar(\mu)=\frac{\za\langle\partial A^a\calO\rangle}
{2\zp(\mu)\langle P^a\calO\rangle}\,,
\end{equation}
where we have not yet specified the source $\calO$ (the running mass
should be practically independent of this) nor the precise definition
of the expectation value. The running of a PCAC mass is determined
by the running of the renormalization constant of the pseudoscalar density 
$\zp$. 

One scheme for computing scale dependent renormalization constants e.g. $\zp$,
is the regularization independent (RI) MOM scheme described
for couplings in sect.~3.1. Originally introduced by  
\shortcite{Martinelli:1993ij}, it is now quite popular 
and used by many collaborations.
Progress using this scheme was reported by Y.~Aoki at LATT09 
\shortcite{Aoki:2009ka}, \shortcite{Sturm:2009kb}
and will be covered in these lectures by Vladikas.

Again an alternative is to apply
finite size recursion techniques here too, and defining $\zp$
running with the volume. Then in the continuum we can define an
associated step scaling function $\sigma_{\rm P}$ through
\be
\frac{\mbar(1/L)}{\mbar(1/2L)}=
\frac{\zp(2L)}{\zp(L)}=\sigma_{\rm P}(u)\,,\,\,\,\,\,
u=\gbar^2(L)\,.
\end{equation}
In perturbation theory we have
\be
\sigma_{\rm P}(u)=1-d_0\ln(2)u+\rmO(u^2)\,.
\end{equation}

After having chosen a definition of $\zp$ on the lattice 
for a given $g_0,L/a$, we can proceed to compute $\sigma_{\rm P}(u)$
via
\ba
\sigma_{\rm P}(u)&=&\lim_{a\to0}\Sigma_{\rm P}(u,a/L)\,,
\\
\Sigma_{\rm P}(u,a/L)&=&\frac{\zp(g_0,2L/a)}{\zp(g_0,L/a)}
\vert_{\gbar^2(L)=u}\,.
\ea
Suppose we have already determined the step scaling function for the running
coupling (as described in sect.~3) at say 8 points:
\be
L_k=2^{-k}\Lmax\,,\,\,\,\,u_k=\gbar^2(L_k)\,,\,\,\,\,k=1,\dots,8\,.
\end{equation}
Then at each step $k$ we measure $\Sigma_{\rm P}(u_k,a/L)$ for a sequence
of values of $L/a$ and extrapolate these to the continuum limit:
\be
\Sigma_{\rm P}(u_k,a/L)=\sigma_{\rm P}(u_k)+\rmO(a^p)\,,
\end{equation}
where $p=2$ for the $\rmO(a)$ improved theory (up to boundary terms).
Thereby we have 
\be
\frac{M}{\mbar(1/\Lmax)}=
\frac{M}{\mbar(1/L_8)}\prod_{k=1}^8\sigma_{\rm P}(u_k)\,.
\end{equation}
Since $L_8$ is a very small physical length we can safely
use PT to determine the first factor $\frac{M}{\mbar(1/L_8)}$.
To relate the RGI mass $M$ to a low energy scale it then remains
to compute $\Lmax$ and $\mbar(1/\Lmax)$ in physical units e.g. $f_\pi$.

\section{The Alpha Collaboration project}

It has been an Alpha Collaboration goal since $\sim 1998$ to measure
running couplings, running quark masses and renormalization constants
in QCD to a high precision. The program uses $\rmO(a)$ improved
Wilson fermions in the SF framework. As we have tried to emphasize
this procedure is comparatively clean and care is taken to control
systematic errors at each stage, but it involves a lot of preparatory
work and attention to details.

Firstly there are extra improvement coefficients needed to cancel
$\rmO(a)$ boundary effects, even for the pure gauge theory
$\sim\int_{\partial\Lambda}F_{0k}^2\,,\int_{\partial\Lambda}F_{ij}^2$.
This involves adjusting weights for the plaquettes at the boundary
\be
w(p)=\left\{\begin{array}{c}
\frac12c_{\rm s}(g_0^2)\,,\,\,{\rm for}\,\,p\in\Lambda\\
c_{\rm t}(g_0^2)\,\, {\rm for}\,\,p\,\,{\rm temporal\,\,touching}
\,\,\Lambda\\
\end{array}
\right\} \,,
\end{equation}
with $c_{\rm s,t}=1+\rmO(g_0^2)$. There are similar coefficients
$\tilde{c}_{\rm s},\tilde{c}_{\rm t}$ for fermionic boundary terms.
Moreover there is a $b_\xi$-coefficient for the renormalized boundary
operator $\xi_{\rm R}=Z_\xi(\widetilde{g}_0^2,a\mu)(1+b_\xi a\mq)$
although these factors usually cancel in ratios defining observables of 
interest. It has been observed that these coefficients have small
coefficients, and thus in practice it is considered safe to use 
the 1-loop perturbative approximation
instead of determining them non-perturbatively. 

Many of the steps must be carried out in a definite order. For example the
SF running coupling is defined as before and at zero quark masses $\mq=0$. 
This means that before measuring the coupling 
the critical mass $\mc$ must be
determined but this in turn depends on $\csw$.
In this project $\mc$ is defined through the vanishing of the
PCAC mass - other definitions e.g. by vanishing pion mass will differ
by $\rmO(a^2)$ in the improved theory.

In an improved theory we expect \shortcite{Luscher:1996ug}
\be
\langle(A_\rmR)^a_\mu(x)\mathcal{J}^a\rangle=
2\mr\langle P_\rmR^a(x)\mathcal{J}^a\rangle+\rmO(a^2)
\end{equation}
for arbitrary sources $\mathcal{J}^a$. In the framework of the SF
we can take $\mathcal{J}^a=\calO^a$ as defined in (\ref{SFsource}).
In terms of the bare correlation functions
\ba
f_A(x_0)&=&-\frac13\langle A^a_0(x)\mathcal{O}^a\rangle\,,
\\
f_P(x_0)&=&-\frac13\langle P^a(x)\mathcal{O}^a\rangle\,,
\ea
we then define improved bare PCAC masses by
\be
m(x_0,\theta_k,C,C')=\frac{\frac12(\partial_0+\partial_0^*)f_{A_I}(x_0)}
{2f_p(x_0)}\,.
\end{equation}
Finally renormalized SF PCAC masses are given by
\be
\mbar_{SF}(L)=\frac{(1+\bA a\mq)\za(g_0)}{(1+\bp a\mq)\zp(g_0,L/a)}m\,,
\end{equation}
where to complete the definitions we must specify the arguments of $m$.

It is clear that the coefficients $\ca,\csw$ can be determined by
demanding that $m$ is independent of the sources up to terms $\rmO(a^2)$.
This can be done by appropriately choosing independent configurations of
$x_0,\theta_k,C,C'$. One can proceed as follows: noting that $m(x_0)$
is linear in $\ca$, $m(x_0)=r(x_0)+\ca s(x_0)$ we can form from 
two choices of boundary conditions $(C_1,C'_1),(C_2,C'_2)$ at say 
$\theta_k=0$, a linear combination 
\be
M_1(x_0,y_0)=m_1(x_0)-s_1(x_0)\frac{[m_1(y_0)-m_2(y_0)]}{s_1(y_0)-s_2(y_0)}
\end{equation}
which is independent of $\ca$ and equal to $m_1(x_0)$  up to $\rmO(a^2)$
corrections (in the improved theory). We can now give a condition 
determining $\csw$ e.g.
\be
0=M_1(T/4,3T/4)-M_2(T/4,3T/4)\,\,\,\,\,({\rm for}\,\,M_1(T/2,T/4)=0)\,.
\end{equation}
A condition such as the latter in brackets is necessary to complete
the specification; to ensure that the results are relatively 
insensitive to the precise choice
one should check that the dependence on $m$ is weak for 
$m\simeq\mc$ (which is only known approximately at this stage). 
Once $\csw$ has been determined, 
we can specify a computation of $\ca$ 
e.g. by comparing values of $m$ at $C=C'=0$ and varying $\theta_k$.

The data for $\csw,\ca$ thus obtained are usually fitted in the measured range
to (rational) functions of $g_0$ which incorporate the known 
perturbative coefficients. These functions can now be considered 
as definitions of the improved theory, which can then be used in other
simulations (not necessarily in the SF framework).

It is important to appreciate that $\csw,\ca$  have $\rmO(a)$
ambiguities; there is no way to remove these outside of PT.
Once these coefficients have been specified, we can use the vanishing
of $m_{\rm PCAC}$  for a determination of $\mc$. Again the precise
value depends on all specific definitions involved.

Knowing $\csw,\mc$ we can now compute $\gbar^2_{\rm SF}(L)$
and obtain the step scaling function in the continuum limit.

Next one can compute the pseudoscalar density scaling function 
$\sigma_P$ as described in sect.~5.3 in a given SF scheme e.g.
\be
\zp(g_0,L/a)=\frac{c\sqrt{f_1}}{\fp(T/2)}
\vert_{T=L,m=\mc,C=C'=0,\theta_k=1/2}\,,
\end{equation}
\be
f_1=-\frac{1}{3L^6}\langle\mathcal{O}^{\prime a}\mathcal{O}^a\rangle\,,
\end{equation}
and $c$ is chosen st $\zp|_{g_0=0}=1$.

Finally one can compute the vector isovector current renormalization
constant $\zv$ using the WI (the derivation is left as an exercise)
\be
(1+\bv a\mq)\zv\fv(x_0)=f_1+\rmO(a^2)\,,
\end{equation}
where
\ba
\fv(x_0)&=&\frac{1}{6L^3}\langle (V_I)_0^a(x)\mathcal{O}^a_{\rm ext}\rangle\,,
\\
\mathcal{O}^a_{\rm ext}&=&
\epsilon^{abc}\mathcal{O}^b\mathcal{O}^{\prime c}\,,
\ea
and for $\za$:
\be
(1+\bA a\mq)\za^2 f_{A_IA_I}(x_0,y_0)=f_1+\rmO(a^2)\,,
\end{equation}
where
\be
f_{AA}(x_0,y_0)=\frac{a^6}{L^6}\sum_{\bfx,\bfy}
\epsilon^{abc}\langle A_0^a(x)A_0^b(y)\mathcal{O}^c_{\rm ext}\rangle\,,
\end{equation}
the conditions for determining $\za,\zv$ being at $\mq=0$.
No explicit results are shown here but the program outlined above 
is practically complete for $\Nf=2$ species of dynamical fermions
(\shortciteNP{DellaMorte:2004bc}, \citeyearNP{DellaMorte:2005dy}, 
\citeyearNP{DellaMorte:2005kg}), \shortcite{DellaMorte:2005rd}, 
\shortcite{DellaMorte:2005se},
and for $\Nf=3$ \shortcite{Aoki:2009tf}. 
In particular the SF coupling is measured 
over a wide range of energies \shortcite{DellaMorte:2004bc}
with a result similar in quality to that for $\Nf=0$.
Finally some results for the Alpha program are recently available
for $\Nf=4$ \shortcite{Tekin:2009kq}.

\chapter{Other improved actions}

Many types of improved actions have been used in large scale simulations.
As mentioned before this is also useful for numerically verifying
universality. Often due to the pressure of completing the simulation 
routines before a
new supercomputer is delivered, pragmatic choices of the action to be 
used have to be made. For example the CPPACS collaboration decided to
employ the Iwasaki action \shortcite{Iwasaki:1983ck} which has the form 
(\ref{Simp}) with
$c_0=3.648,c_1=-0.331,c_2=c_3=0$ independent of $g_0$,
but optimized for values of $g_0$ in the range to be simulated 
(e.g. by demanding good rotational 
symmetry properties of the static potential).          

\section{Perfect actions}

An action used by Hasenfratz, Hasenfratz and Niedermayer 
\citeyear{Hasenfratz:2005tt} is based on Wilson's 
renormalization group approach (see Niedermayer's review 
\citeyear{Niedermayer:1998gk}).
The intriguing realization is that in the huge class of lattice actions 
which have the same universal continuum limit there are ``perfect actions''
for which the physical quantities that can be measured have no lattice 
artifacts at all even when the correlation lengths are $\rmO(a)$!

To understand how this comes about
consider an RG transformation in configuration space of a pure gauge 
theory of the form
\be
\rme^{-\beta' A'(V)}=\int[\rmd U]\,\rme^{-\beta\left[ A(U)+T(U,V)\right]}
\end{equation}
with kernel 
\be
T(U,V)=-\frac{\kappa}{N}\sum_{n_B,\mu}
\Re\,\tr\,\left[V(n_B,\mu)Q^\dagger(n_B,\mu)-N^\beta_{\mu}\right]\,,
\label{RGkernel}
\end{equation}
where $Q(n_B,\mu)$ is a sum over products of link variables $U$ on paths 
from $2n_B$ to $2(n_B+\hat{\mu})$ on the original lattice (e.g. a sum over 
staples). For physical quantities which can be measured with action
$\beta'A'$ we will get the same results as those measured with $\beta A$.
The main problem however is that we are now dealing with an infinite 
coupling parameter space.

The situation is simplified for the case of asymptotically free theories
with one relevant direction labeled by $\beta$, with the critical surface
where the correlation length diverges given by $\beta=\infty$.
If we start the RG transformation with $\beta$ very large then we expect 
that $A'$ will have an expansion of the form
\be
A'(V)=A_0(V)+\frac{1}{\beta}A_1(V)+\dots
\end{equation}
(see Fig.~\ref{RGflow}) where 
\be
A_0(V)=\min_U\left[A(U)+T(U,V)\right]\,.
\end{equation}
Performing this minimization with $A_0(U)$ on the critical surface 
$\beta=\infty$ we stay on that surface. Repeating it infinitely many 
times we finally reach the ``fixed point action'' $A^\mathrm{FP}$ associated 
with the kernel $T$. It satisfies the equation
\be
A^\mathrm{FP}(V)=\min_U\left[A^\mathrm{FP}(U)+T(U,V)\right]\,.
\end{equation}
This action is classically perfect; there exist solutions of the classical 
field equations (e.g. instanton configurations) with no lattice artifacts. 
If one starts the RG transformation close to the FP one obtains an RG 
trajectory along which one has very small lattice artifacts even for very 
small correlation lengths.

\begin{figure}
  \psset{unit=5mm}
  \begin{pspicture}(0,-10)(30,10)
    
    \psset{linewidth=1pt}

    \psset{linewidth=1pt} \psline{->}(10,0)(25,0) \psline{->}(10,0)(10,10)
    \psline{->}(10,0)(5,-5) \rput(5,-3.5){\Large $c_1$} \rput(9,9){\Large
      $c_2$} \rput(24,-1){\Large $g^2$}

    \psset{linewidth=2pt} \psline{->}(4,0)(5.9,1.9) \psline{->}(8,4)(6.1,2.1)
    \psline{->}(6,0)(6,1.9) \psline{->}(6,4)(6,2.1)

    \psline[linewidth=2pt,linestyle=dashed]{-}(6,2)(20,2) \rput(21,2){\Large
      FP}

    \pscurve[linewidth=2pt,showpoints=false]{->}(6,2)(7,2.1)(8,2.4)(9,3)(11,4)
    \pscurve[linewidth=2pt,showpoints=false]{->}(11,4)(13,4.5)(15,5)(20,6)
    \rput(21,6){\Large RG}
  \end{pspicture}
\vspace{-3cm}
\caption{RG flow} \label{RGflow}
\end{figure}

The proposal of Hasenfratz and Niedermayer 
\citeyear{Hasenfratz:1993sp} is to simulate the action 
$\beta A^\mathrm{FP}$ for $\beta$ large. In practice it is too time consuming
to determine $A^\mathrm{FP}$ configuration by configuration. Instead one does 
it for a large number of configurations and parameterizes the results with
a relatively small number of loops. Also they tune the parameter $\kappa$ 
in (\ref{RGkernel}) to weaken the the strength of the spread of loops 
in lattice units.

Fermions can be included along similar lines. A very nice feature is that
fermion actions which are obtained by blocking from a continuum action
which has chiral symmetry have Dirac operators which satisfy the 
Ginsparg--Wilson relation \shortcite{Ginsparg:1981bj}
\be
\left\{\calD,\gamma_5\right\}=(1+s)\calD\gamma_5\calD\,,
\label{GWreln}
\end{equation}
(or some mild local generalization thereof), 
where $s$ is a parameter $|s|\le1$. 
Consider for example lattice
fermions $\chi$ obtained from blocking a free massless 
continuum field $\psi$. The lattice Dirac operator can be
obtained by a minimization procedure over classical fields:
\be
\chibar_n\calD_{nn'}\chi_{n'}=\min_{\psibar,\psi}\left\{\psibar D\psi
+\left(\chibar-\psibar\omega^\dagger\right)
\left(\chi-\omega\psi\right)\right\}\,,
\end{equation}
where
\be
D_{xx'}=\left(\gamma_\mu\partial_\mu\right)_{xx'}\,,
\end{equation}
and
\ba
\omega_{nx}&=&1\,\,\,\mathrm{if}\,\,\,x\in\,\,\mathrm{block}\,\,n
\nonumber\\
&=&0\,\,\,\,\,\,\,\mathrm{otherwise}\,.
\ea
The minimizing field is given by
\be
\psi_0(\chi)=A^{-1}\omega^\dagger\chi\,,\,\,\,\,
\psibar_0(\chi)=\chibar\omega A^{-1}\,,\,\,\,\,
\end{equation}
with
\be
A=D+\omega^\dagger\omega\,.
\end{equation}
\begin{exercise}
Show that the lattice Dirac operator
\be
\calD=1-\omega A^{-1}\omega^\dagger
\end{equation}
satisfies the GW relation (\ref{GWreln}) with $s=1$.
\end{exercise}

Moreover Hasenfratz et al show in a very illuminating paper 
\citeyear{Hasenfratz:2006kv} 
the general result that lattice actions induced by a RG 
procedure inherit all the symmetries of the continuum theory, 
and they give a general procedure 
which delivers the corresponding symmetry transformation on the lattice
e.g. for the U(1) axial continuum transformation
\be
\delta\psi=i\epsilon\gamma_5\psi\,,\,\,\,\,
\delta\psibar=i\epsilon\psibar\gamma_5\,,\,\,\,\,
\end{equation}
one obtains L\"{u}scher's lattice transformation \shortcite{Luscher:1998pqa}
\be
\delta\chi=i\epsilon\gamma_5(1-\calD)\chi\,,\,\,\,\,
\delta\chibar=i\epsilon\chibar(1-\calD)\gamma_5\,.
\end{equation}

\section{Neuberger's action}

There are also classes of Dirac operators satisfying the GW relation which 
are not obtained by RG considerations e.g. Neuberger's overlap massless 
Dirac operator (\shortciteNP{Neuberger:1997fp}, \citeyearNP{Neuberger:1998wv})
\ba
D_N&=&\frac{(1+s)}{a}\left[1-A\left(A^\dagger A\right)^{-1/2}\right]\,,
\\
A&=&1+s-aD_W\,,
\ea
which is discussed in Kaplan's lectures 
(see also Niedermayer's review \citeyear{Niedermayer:1998bi}). 
Simulations with associated actions are highly desirable because of their 
excellent chiral properties, however at present they are very 
(CPU) expensive and hence only pursued by a minority of groups such as 
JLQCD \shortcite{Ohki:2009mt}. 
Here I only want to summarize a few points concerning 
their renormalization and their lattice artifacts.

The Feynman rules are straightforward to derive but a bit more complicated 
than other actions because the action is not ultra-local. The
free massless propagator is rather simple and takes the form
($\ring{p}_\mu=a^{-1}\sin(ap_\mu)$)
\ba
S_0(p)&=&\frac{1}{2(1+s)}\left\{-\frac{i\gamma\ring{p}}{\ring{p}^2}
[\omega(p)+b(p)]+1\right\}\,,
\\
b(p)&=&\frac{(1+s)}{a}-\frac{a}{2}\hat{p}^2\,,\,\,\,
\omega(p)=\sqrt{\ring{p}^2+b(p)^2}\,.
\ea
The quark--antiquark 1-gluon vertex $V(p,q)$ has a relatively 
familiar structure,
however the quark--antiquark 2-gluon vertex involves also an integral over
the product of two such vertices $\sim\int_r V(r,p)V(r,q)K(p,q,r)$
with some kernel $K$. With such a structure it is not immediately clear that
the Reisz conditions for power-counting are met,
however Reisz and Rothe \citeyear{Reisz:1999ck} have proven the 
renormalizability of Neuberger fermions with 2 species 
\footnote{their proof uses the fact that for $\Nf\ge2$ GW lattice fermions 
possess an exact chiral (flavor mixing) symmetry.}
of massless quarks.

As explained in Kaplan's talk the massless GW action can be decomposed 
in a sum of left and right-handed parts
\be
\psibar\calD\psi =\psibar_L\calD\psi_L+\psibar_R\calD\psi_R\,,
\end{equation}
with
\be
\psi_{R/L}=\widehat{P}_\pm\psi\,,\,\,\,\,\,\psibar_{R/L}=\psibar P_\mp\,,
\end{equation}
where the projectors are given by
\ba
P_\pm&=&\frac12\left(1\pm\gamma_5\right)\,,\,\,\,
\widehat{P}_\pm=\frac12\left(1\pm\widehat{\gamma}_5\right)\,,\,\,\,
\\
\widehat{\gamma}_5&=&\gamma_5(1-a\calD)\,.
\ea
It follows that scalar and pseudo-scalar densities are naturally defined by
\ba
S&=&\psibar_L\psi_R+\psibar_R\psi_L=\psibar\left(1-\frac{a}{2}\calD\right)\psi\,,
\\
P&=&\psibar_L\psi_R-\psibar_R\psi_L=\psibar\gamma_5\left(1-\frac{a}{2}\calD\right)
\psi\,,
\ea
which transform into one another under the chiral transformation
\be
\delta\psi_L=-\psi_L\,,\,\,\,
\delta\psibar_L=\psibar_L\,,\,\,\,
\delta\psi_R=0=\delta\psibar_R\,.
\end{equation}
Mass terms are naturally introduced into GW actions via
\be
\calD\rightarrow\calD+mS\,.
\end{equation}
The mass is renormalized multiplicatively as in the continuum.
The pattern of renormalization of quark bilinears and 4--quark operators
is also the same as in the continuum, and for Neuberger fermions computed to 
1--loop perturbation theory by Capitani and Giusti \citeyear{Capitani:2000da}.

Exactly conserved flavor vector currents which do not require 
renormalization have been constructed 
\shortcite{Chandrasekharan:1998wg}, \shortcite{Kikukawa:1998py}, 
\shortcite{Hasenfratz:2002rp}. 
Since these are relatively complicated other currents
of the form $\psibar\gamma_\mu\frac12\left(1-\frac{a}{2}\calD\right)\psi$
are usually used as observables. They also have nice chiral transformation  
properties but are not exactly conserved and hence require finite
renormalization \shortcite{Alexandrou:2000kj}.

The discussions of lattice artifacts are again based on
accepting the validity of arguments based on Symanzik's effective action.
Firstly for $m=0$ there are no $\rmO(a)$ artifacts for spectral 
quantities since in this case any such effects 
would be described by operators of dimension 5 having symmetries of the 
lattice action. No such operator can be constructed for GW fermions 
(a SW-like term breaks the chiral symmetry). When $m\ne0$ the GW 
action is invariant under the transformation
$\psi\to\hat{\gamma}_5\psi,\psibar\to-\psibar\gamma_5$ together 
with $m\to -m$. This symmetry is however not quite sufficient 
to disqualify $\rmO(am)$ artifacts in correlation functions since 
these transformations change the measure for topological 
non-trivial configurations. One can argue however that in large volumes one 
can, for spectral quantities, restrict attention to the topologically 
trivial sector and hence also expect no $\rmO(am)$ artifacts.

Matrix elements of composite operators mentioned above do however have 
$\rmO(a)$ effects. The improvement of bilinears
are obtained by the prescription ($m=0$) 
(\shortciteNP{Capitani:1999uz}, \citeyearNP{Capitani:1999ay}):
\be
(\psibar\Gamma\psi)_I=\psibar\left(1-\frac{a}{2}D\right)\Gamma
\left(1-\frac{a}{2}D\right)\psi\,.
\end{equation}

As a last point I would like to mention that Schr\"{o}dinger functionals 
for overlap fermions have been constructed by L\"{u}scher 
\citeyear{Luscher:2006df}
and in a different way by Sint (for $\Nf$ even) \citeyear{Sint:2007zz}. 
In order to obtain the natural SF boundary conditions (\ref{SFfermionbc})
in the continuum limit,
L\"{u}scher simply modifies the Neuberger operator 
so that it satisfies the GW relation up to terms on the boundary
\be 
\left\{\gamma_5,\calD\right\}=(1+s)a\calD\gamma_5\calD+\triangle_B
\end{equation}
where $\triangle_B\psi(x)=0\,,$ unless $x_0=0,T$. 
A simple proposal is
\be
\calD=\frac{(1+s)}{a}\left[1-\frac12\left(U+\gamma_5 
U^\dagger\gamma_5\right)\right]\,,
\end{equation}
with
\ba
U&=&A\left(A^\dagger A+caP\right)^{-1/2}\,,
\\
P\psi(x)&=&\frac{1}{a}\left\{\delta_{x_0,a}P_-\psi(x)|_{x_0=a}
+\delta_{x_0,T-a}P_+\psi(x)|_{x_0=T-a}\right\}\,,
\ea
where $c$ is a parameter which can be tuned for $\rmO(a)$ improvement,
and has been computed to 1-loop order PT by Takeda \citeyear{Takeda:2008ni}.

\section{Twisted mass lattice QCD}

As the last topic I would like to mention Wilson twisted mass lattice QCD;
and refer the reader to excellent reviews by Sint \citeyear{Sint:2007ug}
and Shindler \citeyear{Shindler:2007vp}.
For the case of two flavors ($\Nf=2$) the action takes the form
\ba
S_{\rm TM}&=&\int\rmd\chi\rmd\chibar\, \chibar D_{\rm TM}\chi\,,
\\ 
D_{\rm TM}&=&D_W+m_0+i\muq\gamma_5\tau^3\,.
\ea
It appeared first in papers by Aoki and Gosch 
(\citeyearNP{Aoki:1989rw}, \citeyearNP{Aoki:1990ap})
who were considering the extra twisted mass term as an external probe to
study the phase diagram of Wilson fermions. In these works it was realized
that the associated hermitian operator $Q\equiv\gamma_5 D_{TM}$ has no 
zero modes for $\muq>0$:
\be
Q=Q_W+i\muq\tau^3\,,\,\,\,\mathrm{and}\,\,\,Q_W=Q^\dagger_W
\,\,\,\Rightarrow Q^\dagger Q=Q_WQ^\dagger_W+\muq^2\,.
\end{equation}
Thus adding a twisted mass term provides a local field theoretic solution 
to the problem of treating exceptional configurations
\footnote{These are configurations which have exceptionally small 
eigenvalues of $Q_W$ and hence give huge contributions to some correlation 
functions e.g. $\langle P^a(x)P^a(y)\rangle$ in quenched simulations.} 
in quenched simulations. The TM formulation also has some advantages in 
dynamical HMC simulations.

Soon it was realized 
(\shortciteNP{Frezzotti:2000nk}, \shortciteNP{Frezzotti:2001ea})
that TMQCD is a viable alternative regularization of QCD, 
and it is, for an additional reason that I will soon explain, now used in 
some large scale simulations, This may at first sight seem strange 
because parity is apparently strongly violated, but this is not the case.
To see this consider first TMQCD in the formal continuum limit
\be
S_{\rm TMQCD}=\int\chibar\left(\gamma D+\mq+i\muq\gamma_5\tau^3\right)\chi\,.
\end{equation}
Now making a change of variables in the functional integral
\ba
\psi&=&V(\omega)\chi\,,\,\,\,\,
\psibar=\chibar V(\omega)\,,
\\
V(\omega)&=&\exp\left(i\omega\gamma_5\frac{\tau^3}{2}\right)\,,
\,\,\,\,\mathrm{with}\,\,\,\,\tan\omega=\frac{\muq}{\mq}\,.
\ea
the action takes the usual form
\be
S_{\rm TMQCD}=\int\psibar\left(\gamma D+M\right)\psi\,,
\end{equation}
with
\be
M=\sqrt{\mq^2+\muq^2}\,.
\end{equation}
Hence continuum twisted mass QCD is formally equivalent to QCD. The symmetries 
in one basis are in one-to-one correspondence with symmetries in the 
other basis. The $\psi$'s are called the ``physical basis'' because 
in this basis the physical interpretation is clearer and $\chi$'s are
called the TM basis. 

The formal equivalence can be made more rigorous \shortcite{Frezzotti:2000nk}
by considering a lattice regularization of TMQCD which preserves 
a chiral symmetry such as GW fermions. In such a case the 
regularized correlation functions in the different bases are equal:
\be
\langle\calO[\psi,\psibar]\rangle_{(M,0)}=
\langle\calO[\chi,\chibar]\rangle_{(\mq,\muq)}\,.
\end{equation}
Operators in one basis are equivalent to associated operators in the other;
this requires some familiarization e.g. for fermion bilinears in the
TM basis define
\be
V^a_\mu=\chibar\gamma_\mu\frac{\tau^a}{2}\chi\,,\,\,\,
\mathrm{similarly}\,\,\,A^a_\mu,S^a,P^a\,\,\,\mathrm{and}\,\,\,
S^0=\chibar\chi\,,
\end{equation}
and in the physical basis
\be
\calV^a_\mu=\psibar\gamma_\mu\frac{\tau^a}{2}\psi\,,\,\,\,
\mathrm{similarly}\,\,\,\calA^a_\mu,\calS^a,\calP^a\,\,\,
\mathrm{and}\,\,\,\calS^0=\psibar\psi\,.
\end{equation}
Then these are related by
\ba
\calA^a_\mu&=&\cos\omega\,A^a_\mu+\epsilon^{3ab}\sin\omega\, 
V^a_\mu\,,\,\,\,a=1,2\,,
\\
&=&A^a_\mu\,,\,\,\,a=3\,,
\ea
and similarly for $\calV^a_\mu$, and
\ba
\calP^a&=&P^a\,,\,\,\,a=1,2
\\
&=&\cos\omega\,P^3+2i\sin\omega\,S^0\,,\,\,\,a=3\,,
\\
\calS^0&=&\cos\omega S^0+2i\sin\omega P^3\,.
\ea
The familiar PCVC and PCAC relations
\ba
\partial_\mu\calV^a_\mu&=&0\,,
\\
\partial_\mu\calA^a_\mu&=&2M\calP^a\,,
\ea
translate in the TM basis to
\ba
\partial_\mu V^a_\mu&=&-2\muq\epsilon^{3ab}P^b\,,
\\
\partial_\mu A^a_\mu&=&2\mq P^a+i\muq\delta^{3a}S^0\,.
\ea
For the case of a regularization which breaks chiral symmetry, such as 
Wilson TM fermions, one expects that near the continuum limit 
\be
\langle\calO_\rmR[\chi,\chibar]\rangle_{(\mq,\muq)}=
\langle\calO_\rmR[\psi,\psibar]\rangle^\mathrm{cont}_{(M,0)}+\rmO(a)\,.
\end{equation}
The lattice artifacts depend on $\mq$ and on the twist angle $\omega$.
In this sense ordinary Wilson fermions are just a special case of the
WTM regularization of QCD.

To discuss the lattice artifacts the
determination of the structure of the Symanzik effective action again 
plays a central role. This first requires  
listing the symmetries of the WTM action; among these are

\begin{enumerate}

\item gauge invariance, lattice rotations and translations, and charge
conjugation

\item $\calP_F^a$--symmetry which is conventional parity $\calP$  
combined with a flavor rotation:

$\calP_F^a: \chi(x)\rightarrow i\tau^a\calP \chi(x)\,,\,\,\,\,a=1,2\,\,\,\,
\calP\chi(x_0,\bfx)=\gamma_0\chi(x_0,-\bfx)$

\item and $\widetilde{\calP}\equiv\calP\times\{\muq\to-\muq\}$

\item similarly for time reversal

\item also the $\mathrm{U}_V(1)$ subgroup of $\mathrm{SU}_V(2)$ is maintained 

\end{enumerate}

With this knowledge one can again show that the only terms in the on-shell 
effective action (apart from ones that can be incorporated in 
the renormalizations of the bare parameters) is the SW term
\be
S_1^\mathrm{eff}\propto
\int\rmd^4x\,\chibar(x)\sigma_{\mu\nu}F^{\mu\nu}(x)\chi(x)\,.
\end{equation}
The $\rmO(a)$ improvement program of the Wilson 
fermions can be extended 
to the case of arbitrary twist angle $\omega\ne0$\,. In general more 
$b$--coefficients are required e.g. for the mass independent 
renormalization one has to introduce improved bare mass parameters
\ba
\widetilde{m}_\mathrm{q}&=&\mq(1+\bm a\mq)+\tilde{b}_m a\muq^2\,,
\\ 
\widetilde{\mu}_\mathrm{q}&=&\muq(1+\bmu a\mq)\,,
\ea
and for the renormalized axial current
\be
(A_\rmR)^a_\mu=\za(1+\bA a\mq)[A_\mu^a+a\ca\partial_\mu 
P^a+a\muq\tilde{b}_A\epsilon^{3ab}V_\mu^b\,.
\end{equation}

An obvious simplification of the regularization at maximal twist angle
$\omega=\pi/2$ was realized early on, but TMQCD only became more popular 
as a practical regularization after the realization 
by Frezzotti and Rossi \citeyear{Frezzotti:2003ni} 
of a related property called ``automatic O($a$) improvement''. This is
the property that properly renormalized physical expectation values have 
only $\rmO(a^2)$ artifacts provided that the untwisted mass is tuned to
its critical value, e.g. by tuning the PCAC mass 
$\propto\langle\partial_\mu A_\mu^1\calO\rangle$ to 0.
The original demonstration \shortcite{Frezzotti:2003ni} of this was 
not complete, 
but the discoverers and many other authors developed independent proofs
\shortcite{Sharpe:2004ny}, \shortcite{Shindler:2005vj}, 
\shortcite{Sint:2005qz}, (\shortciteNP{Aoki:2004ta}, 
\citeyearNP{Aoki:2006nv}), \shortcite{Frezzotti:2005gi}.

Working in the TM basis with
\ba
S^\mathrm{eff}&=&S_0+aS_1^\mathrm{eff}\,,
\\
S_0&=&\int\rmd^4x\,\chibar\left[\gamma D+i\muq\gamma_5\tau^3\right]\chi
\ea
we have
\be
\langle\calO_\rmR\rangle=\langle\calO_R\rangle^\mathrm{cont}
-a\langle\calO_\rmR S_1^\mathrm{eff}\rangle^\mathrm{cont}
+a\langle\calO_1^\mathrm{eff} \rangle^\mathrm{cont}+\rmO(a^2)\,.
\end{equation}
One proof is based on the following symmetry of $S_0$:
\be
\calR^1_5:\,\,\,\,\chi\to i\tau^1\gamma_5\chi\,,\,\,\,\,\,
\chibar\to \chibar i\tau^1\gamma_5\,.
\end{equation}
$\calR^1_5$ alone is not a symmetry of WTM, but $\calR^1_5\times\calD$ is,
where $\calD$ essentially measures the parity of the dimensions of 
operators
\ba
\calD:\,\,\,\,\,U(x,\mu)&\to &U^\dagger(-x-a\hat{\mu},\mu)\,,
\\
\chi(x)&\to &\rme^{3i\pi/2}\chi(-x)\,,\,\,\,
\chibar(x)\to \rme^{3i\pi/2}\chibar(-x)\,.
\ea
Consider now operators $\calO_\pm$ which are even(odd) under $\calR^1_5$.
Now $S_1^\mathrm{eff}$ is odd under $\calR^1_5$; also (using $\calD$)
the effective operators $\rmO_{\pm1}^\mathrm{eff}$ are odd(even) under 
$\calR^1_5$, and so it follows 
\ba
\langle\calO_{+R}\rangle&=&\langle\calO_{+R}\rangle^\mathrm{cont}+\rmO(a^2)\,,
\\
\langle\calO_{-R}\rangle&=&\rmO(a)\,.
\ea

One can extend the TM regularization to include non-degenerate quarks
and also further flavors in various ways.

Another advantage of using the TM formulation is in the simplification 
of some renormalization properties of composite operators. For example
in the computation of $f_\pi$ with Wilson fermions from the 
correlator $\langle\calA_{\rmR 0}^1(x)\calP_\rmR^1(y)\rangle$ one needs $\za$,
but using the equivalence
\be
\langle\calA_{\rmR 0}^1(x)\calP_\rmR^1(y)\rangle_{(M_\rmR,0)}
\simeq \langle V_0^2(x)P_\rmR^1(y)\rangle_{(M_\rmR,\pi/2)}
\end{equation}
no renormalization constant for the current on the rhs is needed.

Another example is the 4-fermion operator appearing in the description of
$\triangle S=2$ transitions. In the physical basis
\be
\calO^{\triangle S=2}_{(\calV-\calA)(\calV-\calA)}
=\sum_\mu\left(\bar{s}\gamma_\mu(1-\gamma_5)d\right)^2
\end{equation}
is a sum of two parts, one parity even  $\calO_{\calV\calV+\calA\calA}$
and the other parity odd  $-\calO_{\calV\calA+\calA\calV}$. As mentioned in 
sect.~2.7 for Wilson fermions the parity even operator mixes with many
other operators. However with WTM regularization at maximal 
twist in the $u,d$ sector and Wilson for the $s$--quarks one obtains
\be
\calO_{\calV\calV+\calA\calA}\simeq\calO_{VA+AV}\,.
\end{equation}
Using $C,\widetilde{P}$ symmetry one can show that $\calO_{VA+AV}$
renormalizes diagonally, which simplifies the problem considerably
\shortcite{Pena:2004gb}.

In conclusion, TMQCD is a competitive practical regularization of QCD,
especially at maximal twist. The price to be paid however is a breaking of 
parity and flavor symmetries. This is usually at the $\rmO(a^2)$ level but
these effects can in some situations be uncomfortably large. 


\end{document}